\documentclass[epj,nopacs,iicol]{svjour}
\usepackage{latexsym}
\usepackage{multirow}
\usepackage{algpseudocode}
\usepackage{algorithm}
\usepackage{graphicx}
\usepackage{amsmath}
\usepackage{amssymb}
\usepackage{amsfonts}
\usepackage{times}
\usepackage{xspace} 
\usepackage[usenames]{color}
\usepackage{dcolumn}
\usepackage{bm}
\usepackage{mathrsfs}
\usepackage{tikz}
\usepackage{appendix}
\usepackage{dsfont}
\usepackage{breqn}
\usepackage[]{units}       % for units environment
\usepackage{epstopdf}

\usepackage{subfigure}		% for subfigures
\usepackage[normalem]{ulem} % for strikeout

\newcommand{\take}[2]{\begin{pmatrix}
#1 \\ #2
\end{pmatrix}}
\newcommand{\deri}[2]{\dfrac{\textnormal{d} #1 }{\textnormal{d} #2}}

\emergencystretch=\maxdimen
\makeatletter
\let\cat@comma@active\@empty
\makeatother
\begin{document}
\title{Bernstein spectral method for quasinormal modes and other eigenvalue problems}
\author{Sean Fortuna \inst{1} \and Ian Vega\inst{1}% etc
}                     % Do not remove
\offprints{}          % Insert a name or remove this line
\institute{National Institute of Physics, University of the Philippines Diliman, 75 C.P. Garcia Ave, Quezon City, 1101, Philippines}
\date{Received: 2023 / Revised version: }
\abstract{
Spectral methods are now common in the solution of ordinary differential eigenvalue problems in a wide variety of fields, such as in the computation of black hole quasinormal modes. Most of these spectral codes are based on standard Chebyshev, Fourier, or some other orthogonal basis functions. In this work we highlight the usefulness of a relatively unknown set of non-orthogonal basis functions, known as Bernstein polynomials, and their advantages for handling boundary conditions in ordinary differential eigenvalue problems. We also report on a new user-friendly package, called \texttt{SpectralBP}, that implements Berstein-polynomial-based pseudospectral routines for eigenvalue problems. We demonstrate the functionalities of the package by applying it to a number of model problems in quantum mechanics and to the problem of computing scalar and gravitational quasinormal modes in a Schwarzschild background.  We validate our code against some known results and achieve excellent agreement. Compared to continued-fraction or series methods, global approximation methods are particularly well-suited for computing purely imaginary modes such as the algebraically special modes for Schwarzschild gravitational perturbations.
\PACS{
      {PACS-key}{discribing text of that key}   \and
      {PACS-key}{discribing text of that key}
     } % end of PACS codes
} %end of abstract
\maketitle
\section{Introduction}
\label{sec:introduction}

Black holes in general relativity are simple spacetime objects, fully specified by only a handful of constants. When the spacetime around black holes is disturbed by surrounding complex distributions of matter and fields, as they are found in nature, these spacetime disturbances generically evolve in the form of damped oscillations known as \textit{quasinormal modes} (QNMs).

Quasinormal modes are the characteristic ringing of spacetime around black holes. They are independent of the initial excitation that generated them, dependent only on parameters of the black hole. A wealth of information can be extracted from the quasinormal mode spectrum of a black hole, so they serve as probes for the validity of general relativity and its extensions in the strong gravity regime. Two excellent reviews on the topic with an emphasis on astrophysics can be found in \cite{Kokkotas1999} and \cite{Nollert1999}. A review on higher dimensional black holes and their connections to strongly coupled quantum fields can be found in \cite{Berti2009}.

In general, the quasinormal mode spectrum of a black hole comes from solving an ordinary differential equation (ODE) eigenvalue problem. These usually take the form of a Schr\"{o}dinger-like equation,
\begin{equation}
-\deri{^2 R}{r_*^2} + V(r, \omega)R = \omega^2 R.
\label{eq:SchrodingerEquation}
\end{equation}

\noindent where $r_*$ is called a tortoise coordinate. 

Various numerical methods have been developed to solve \eqref{eq:SchrodingerEquation}, such as the WKB approach, shooting methods, continued-fraction methods, and the use of P\"{o}schl-Teller potentials. A review article with an emphasis on this topic can be found in \cite{Konoplya2011}. In this paper, we shall be solving \eqref{eq:SchrodingerEquation} using a pseudospectral method.

The use of spectral and pseudospectral methods in gravitational problems is well-established \cite{Grandclement2009,Dias2016}, and have been applied to numerous numerical experiments such as \cite{Cardoso2014,Dias2015,Cayuso2020} to name a few. Here we extend this library of methods to include the Bernstein polynomial basis, which has particular properties that lend its use to mixed-type boundary-value problems.

Likewise, Bernstein polynomials have been used as a function basis in the numerical solution of various differential \cite{Bhatta2006,IdreesBhatti2007,Doha2011,Doha2011a,Tabrizidooz2018}, fractional differential \cite{Yuzbasi2013}, integral \cite{Mandal2007,Maleknejad2012,Maleknejad2012a,Hesameddini2017}, integro-differential \cite{Bhattacharya2008,Yuzbasi2016} and fractional integro-differential \cite{Mirzaee2017} equations. Multiple methods have been deployed in this context, such as the Bernstein-Petrov-Galerkin (BPG) method, the collocation method, operational matrices and direct integration. Our work extends the range of the Bernstein basis by exploring its use in ODE \emph{eigenvalue} problems.

The aim of this paper is two-fold. First, it is a primer on how Bernstein polynomials (BPs) may be used for boundary-value problems in a general relativity setting. Second, it is an introduction to a \texttt{Mathematica} package we call \texttt{SpectralBP} that implements the pseudospectral method based on Bernstein polynomials. For examples and benchmarks, we have applied \texttt{SpectralBP} to a selection of eigenvalue problems in quantum mechanics and general relativity: the infinite square well, harmonic and anharmonic oscillators, and quasinormal modes of various fields in a Schwarzschild black hole. Particularly noteworthy is that \texttt{SpectralBP} is able to find eigenvalues with modest resources where other numerical methods find with difficulty, such as the algebraically special modes for gravitational perturbations of the Schwarzschild geometry \cite{Andersson1994,Leaver2006}. As will be explained below, this should be expected of any spectral method for eigenvalue problems.

The method introduced in this paper, and the accompanying \texttt{Mathematica} package, has seen use in general relativity \cite{sanches2022,konoplya2022d,fortuna2022,konoplya2022a,konoplya2022,konoplya2023,fu2023,konoplya2023a,gogoi2023} and quantum mechanics \cite{Galapon2023}. It has been particularly useful in finding purely imaginary quasinormal modes \cite{konoplya2023a,konoplya2022a}, finding new branches of solutions hitherto unknown \cite{konoplya2022,konoplya2022d} and show novel and critical behaviors like spectrum bifurcation \cite{fortuna2022} and instability \cite{konoplya2023}. In quantum mechanics applications, it has been shown to generate exceedingly accurate solutions where other methods require vast resources in memory and compute time \cite{Galapon2023}.

Consider an $n \times n$ matrix of linear differential operators $\hat{L}(u,\omega)$ dependent on a single independent variable $u$ and polynomial in the eigenvalue $\omega$ of some maximal integer order $m$,
\begin{equation}
\begin{array}{l}
\hat{L}_{i,j} (u,\omega) = \hat{f}_{i,j,0} + \omega \hat{f}_{i,j,1} + \dots + \omega^m \hat{f}_{i,j,m},\vspace{10pt}  \\
\hat{f}_{i,j,k} = f_{i,j,k}(u, \deri{}{u},\deri{^2}{u^2},\dots),
\end{array}
\end{equation}

\noindent and let $\Phi(u)$ be a vector of $n$ functions dependent on $u$
\begin{equation}
\Phi(u) = (\phi_1 (u), \phi_2 (u), \dots, \phi_n(u))^T.
\end{equation}

\noindent We wish to solve the following eigenvalue problem for $\omega$,
\begin{equation}
\hat{L}(u,\omega) \Phi(u) = 0,
\label{eq:fullproblem}
\end{equation}

\noindent provided the problem satisfies the following criteria:
\begin{enumerate}
\item The domain of the solution is compact and analytic over its whole domain. ($u \in [a,b]$)
\item The boundary conditions for all eigenfunctions $\psi_i(u)$ specifies that $\lim_{u \to a} \psi_i(u) \sim (u-a)^{q}$ and $\lim_{u \to b} \psi_i(u) \sim (b-u)^{r}$ for some $q, r \geq 0$.
\item The eigenvalues of $\omega$ form a discrete spectrum.
\end{enumerate}

The calculation of the bound state energies of quantum mechanical particles and the quasinormal modes of black hole spacetimes are examples of such a problem.

To solve \eqref{eq:fullproblem} we use a pseudospectral method, in which the solution of the differential equation is approximated as a weighted sum of a set of basis functions, say $\{\phi_i(r)\}$, as in,
\begin{equation}
R(r) \approx \sum_i C_i \phi_i(r).
\end{equation}
This renders the initial differential problem into a system of algebraic equations the set of expansion coefficients $\{C_i\}$ must satisfy. Since \eqref{eq:fullproblem} is linear, these algebraic equations can be cast as a matrix equation generically of the form of a generalized eigenvalue problem (GEP),
\begin{equation}
\bm{M}(\omega) \bm{C} = 0.
\label{eq:GenericMatrixEquation}
\end{equation}

We have developed a \texttt{Mathematica} package we call \texttt{SpectralBP}, written to streamline the numerical solution of ODE eigenvalue problems. The package utilizes the Bernstein polynomials, and the properties which make them particularly powerful in the context of boundary value problems.

A similar \texttt{Mathematica} package can be found in \cite{Jansen2017}. It is a pseudospectral method which uses a Chebyschev polynomial basis, called \texttt{QNMSpectral}. This open-source package served as the initial inspiration for our work, and so the two codes unavoidably overlap in some of their functionality. We developed \texttt{SpectralBP} to be a superset of \texttt{QNMSpectral}, with the intent of developing a spectral solver not just specifically tailored to quasinormal mode calculations. It also serves to introduce the Bernstein method to the general relativity community. Aside from methods specifically tied to the Bernstein basis, \texttt{SpectralBP} also implements a novel algorithm for efficiently tackling transcendental and polynomial eigenvalue problems that we shall discuss in detail in a future paper \cite{fortuna2020c}.

This paper is organized in two parts. We first establish how the Bernstein polynomial basis may be used in ODE eigenvalue problems with boundary conditions. In Section~\ref{sec:preliminaries}, we fix our notation and enumerate the properties of the Bernstein basis relevant to the method. In Section~\ref{sec:boundary}, we explain how the Bernstein basis is appropriate in handling boundary conditions. In Section~\ref{sec:pseudospectral}, we review standard methods for translating 
\eqref{eq:fullproblem} into a generalized eigenvalue problem using a collocation method. We then enumerate various positives and negatives the Bernstein polynomial basis has compared to other bases like Fourier or Chebyschev in Section~\ref{sec:procon}.

The rest of the paper involves the implementation and application of \texttt{SpectralBP}. Section~\ref{sec:SBPintro} introduces the \texttt{SpectralBP} package and its general features. We then show in detail how \texttt{SpectralBP} can be used in Section \ref{sec:TISE} and Section \ref{sec:QNM}, introducing functionalities of the package by working out some model problems in quantum mechanics and calculating quasinormal modes respectively. In Section~\ref{sec:ASM}, we look at the algebraically special modes of the Regge-Wheeler equation. In the final section, we show miscellaneous details implemented in \texttt{SpectralBP}: closed-form expressions of the spectral matrices, matrix inversion, and eigenfunction calculation and manipulation.

\section{Bernstein polynomials}
\label{sec:preliminaries}

We review some of the key properties of Bernstein polynomials. We shall not be exhaustive and select only those properties useful to the development of \texttt{SpectralBP}. This section shall also fix our notation for the rest of the paper. A useful reference can be found in \cite{Doha2011}, which describes all of the properties listed here using a Bernstein basis over the interval $[0,1]$. The generalization to a Bernstein basis over an arbitrary interval $[a,b]$ is straightforward.

The Bernstein basis of degree $N$ defined over the interval $u \in [a,b]$ is a set of $N+1$ polynomials, $\{B^N_k(u)\}$, given by
\begin{equation}
\begin{array}{l}
B^N_k(u) = \take{N}{k} \dfrac{(u-a)^k (b-u)^{N-k}}{(b-a)^N},\vspace{10pt} \\
k = 0,1,\dots,N, \qquad \take{N}{k} = \dfrac{N!}{(k)!(N-k)!}.
\end{array}
\label{eq:BPdefinition}
\end{equation}
For convenience, we also set $B^N_k(u) = 0$ and $\take{N}{k} = 0$ when either $k < 0$ or $k > N$.
\begin{figure}[t]
\begin{center}
\includegraphics[width=0.4\textwidth]{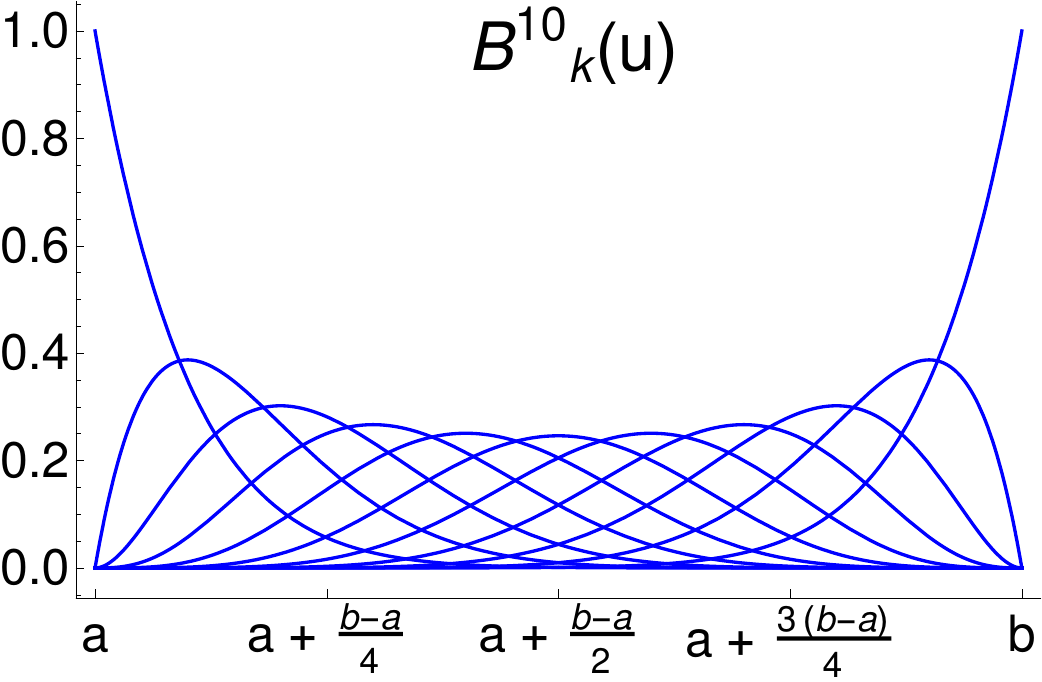}
\caption{Bernstein polynomials of degree 10.}
\label{fig:BP10}
\end{center}
\end{figure}

The Bernstein basis of degree 10 is shown in Figure \ref{fig:BP10}. It is clear that at the boundaries $u=a$ and $u=b$, Bernstein polynomials satisfy 
\begin{equation}
B^N_k(a) = \delta_{k,0}, \qquad B^N_k(b) = \delta_{k,N}.
\label{eq:boundary}
\end{equation}
The derivative of a Bernstein polynomial of degree $N$ can be expressed in terms of Bernstein polynomials of degree $N-1$, satisfying the following recurrence relation,
\begin{equation}
\deri{B^N_k}{u}  = \dfrac{N}{b-a} \left( B^{N-1}_{k-1}(u) - B^{N-1}_{k}(u) \right).
\end{equation}
Repeated differentiation also gives
\begin{multline}
\deri{^m B^N_k}{u^m}  = \dfrac{1}{(b-a)^m} \dfrac{N!}{(N-m)!} \times \\ \sum_{l=0}^m (-1)^l \take{m}{l} B^{N-m}_{k+l-m}(u).
\label{eq:BPderivative}
\end{multline}
%\begin{equation}
%\deri{^m B^N_k}{u^m}  = \dfrac{1}{(b-a)^m} \dfrac{N!}{(N-m)!} \sum_{l=0}^m (-1)^l \take{m}{l} B^{N-m}_{k+l-m}(u).
%\label{eq:BPderivative}
%\end{equation}
A Bernstein polynomial of degree $N$ can be expressed as a sum of Bernstein polynomials of a higher degree \cite{Farouki2000},
\begin{equation}
B^N_k(u) = \sum_{j=0}^{m} \dfrac{\take{N}{k} \take{m}{j}}{\take{N+m}{k+j}} B^{N+m}_{k+j}(u).
\label{eq:BPraising}
\end{equation}
The integral of each basis polynomial in a Bernstein basis of degree $N$ over $[a,b]$ are equal,
\begin{equation}
\int_a^b B^N_k(u) du = \dfrac{b-a}{N+1}.
\label{eq:BPint}
\end{equation}
Finally, the product between two Bernstein polynomials can be expressed as single Bernstein polynomial of higher degree,
\begin{equation}
B^N_j (u) B^M_k (u) = \frac{\take{N}{j}\take{M}{k}}{\take{N+M}{j+k}} B^{N+M}_{j+k} (u).
\label{eq:BPproduct}
\end{equation}

\section{Boundary conditions}
\label{sec:boundary}

When using the Bernstein basis in mixed-type boundary-value problems, we shall see that the boundary conditions act only on a subset of the Bernstein basis. This lets us independently solve the boundary conditions separately, making the Bernstein basis particularly useful in mixed-type boundary-value problems. For the particular boundary-value problem described in Section~\ref{sec:introduction}, the Bernstein method reduces to a form in which each basis function satisfies the boundary conditions.

We begin by approximating the solution $\phi(u)$ as a weighted sum of Bernstein polynomials,
\begin{equation}
\phi(u) \approx \sum_{k=0}^N C_{k} B^N_k(u).
\label{eq:BPapprox}
\end{equation}
Let there be $q$ boundary conditions on $u=a$ and $r$ boundary conditions on $u=b$ of the following form,
\begin{equation}
\hspace{20pt}
\begin{array}{l}
\phi(a) = a_0, \deri{\phi(a)}{u} = a_1, \dots, \deri{^{q-1} \phi(a)}{u^{q-1}} = a_{q-1},\vspace{6pt} \\
\phi(b) = b_0, \deri{\phi(b)}{u} = b_1, \dots, \deri{^{r-1} \phi(b)}{u^{r-1}} = b_{r-1}.
\end{array}
\label{eq:BoundaryConditions1}
\end{equation}
These constants may be interrelated. A common example would be a two-point boundary value problem of a second-order differential equation subject to mixed linear boundary conditions,
\begin{equation}
\;\;
\begin{array}{r}
c_{1,k} \phi(a) + c_{2,k} \phi'(a) + c_{3,k} \phi(b) + c_{4,k} \phi'(b) = c_{5,k}, \\
(k = 1,2,3,4),
\end{array}
\end{equation}
which fixes $a_0, a_1, b_0,$ and $b_1$.

Combining \eqref{eq:BPderivative} and \eqref{eq:BPapprox}, the $m^{\mathrm{th}}$ derivative of $\phi(u)$ is given by
\begin{equation}
\deri{^m \phi}{u^m}  = \sum_{k=0}^N \sum_{l=0}^m \dfrac{C_{k}}{(b-a)^m} \dfrac{N!}{(N-m)!} (-1)^l \take{m}{l} B^{N-m}_{k+l-m}(u).
\end{equation}
We use \eqref{eq:boundary} to simplify evaluating $\phi(u)$ at the boundaries. At $u=a$ and $u=b$, we get
\begin{equation}
\left. \deri{^m\phi}{u^m} \right\rvert_a = \dfrac{1}{(b-a)^m} \dfrac{N!}{(N-m)!} \sum_{l=0}^m C_{m-l} (-1)^l \take{m}{l}
\end{equation}
and
\begin{equation}
\left. \deri{^m \phi}{u^m} \right\rvert_b = \dfrac{1}{(b-a)^m} \dfrac{N!}{(N-m)!} \sum_{l=0}^m C_{N-l} (-1)^l \take{m}{l}.
\end{equation}
Thus, the boundary conditions act only first $q$ and last $r$ of the Bernstein basis, whose expansion coefficients are fixed via the matrix equations
\begin{equation}
\textbf{A} \textbf{C} = \textbf{a}, \qquad \textbf{B} \tilde{\textbf{C}} = \textbf{b},
\label{eq:tau1}
\end{equation}
where
\begin{equation}
\begin{array}{l}
\textbf{A}_{l,m} = \dfrac{1}{(b-a)^l} \dfrac{N!}{(N-l)!}(-1)^{l-m} \take{l}{l-m},\vspace{6pt}\\
\textbf{C}_m = C_m, \; \textbf{a}_l = a_l,\vspace{6pt} \\
m,l \in \{0,1,\dots,q-1\},
\end{array}
\label{eq:tau2}
\end{equation}
and
\begin{equation}
\begin{array}{l}
\textbf{B}_{l,m} = \dfrac{1}{(b-a)^l} \dfrac{N!}{(N-l)!}(-1)^{m} \take{l}{m},\vspace{6pt}\\
\tilde{\textbf{C}}_m = C_{N-m}, \; \textbf{b}_l = b_l,\vspace{6pt} \\
m,l \in \{0,1,\dots,r-1\}.
\end{array}
\label{eq:tau3}
\end{equation}
When the differential operator is linear, the modified ODE eigenvalue problem
\begin{equation}
\hat{L}(u,\omega)\psi(u) = g(u,\omega), \quad \psi(u) = \sum_{k=q}^{N-r} C_k B^N_k(u) 
\label{eq:ODEwithresidual}
\end{equation}
determines the rest of the expansion coefficients, where the residual function $g(u,\omega)$ is given by
\begin{multline}
g(u,\omega) = -\hat{L}(u,\omega)\left( \sum_{k=0}^{q-1} C_k B^N_k(u) \right. \qquad \qquad \\
\qquad \qquad \left. + \sum_{k=N-r+1}^N C_k B^N_k(u) \right).
\label{eq:ODEresidual}
\end{multline}
%\begin{equation}
%g(u,\omega) = -\hat{L}(u,\omega)\left( \sum_{k=0}^{q-1} C_k B^N_k(u) + \sum_{k=N-r+1}^N C_k B^N_k(u) \right).
%\label{eq:ODEresidual}
%\end{equation}
We consider the case where $g(u,\omega)$ vanishes, or equivalently
\begin{equation}
\lim_{u \to a} \phi(u) \sim (u-a)^q, \qquad \lim_{u \to b} \phi(u) \sim (b-u)^r.
\label{eq:BoundaryConditions2}
\end{equation}
We arrive at an ODE eigenvalue problem identical to the one we started with, but over a smaller set of basis functions
\begin{equation}
\hat{L}(u,\omega)\psi(u) = 0, \quad \psi(u) = \sum_{k=q}^{N-r} C_k B^N_k(u). 
\label{eq:probsol}
\end{equation}
It should be noted that for more standard basis functions, imposing the boundary conditions considered in \eqref{eq:BoundaryConditions1} would involve the entire basis set. To determine the expansion coefficients, the differential equations and the boundary conditions must be solved simultaneously. In the Bernstein basis, the boundary conditions act only on the first $q$ and last $r$ basis polynomials, and we get their corresponding expansion coefficients for free even before considering the ODE. Though we do not prove that this advantage is unique to the Bernstein basis, we believe that any other basis must behave like Bernstein polynomials to enjoy it. That is, the $n$th basis function of a basis of size $N$ must asymptote to $(u-a)^n$ towards the lower boundary and to $(b-u)^{N-n}$ towards the upper boundary. 

We express a similar sentiment for other basis functions where the condition \eqref{eq:BoundaryConditions2} would make the residual function vanish. In the Bernstein basis, the problem is simplified since each basis polynomial satisfies the boundary conditions exactly.

Finally, we note that when the differential operator is not dependent on $\omega$, equation \eqref{eq:ODEwithresidual} serves as a general recipe for solving boundary value problems using Bernstein polynomials. One may modify the many methods found in Section \ref{sec:introduction} to solve for the remaining undetermined coefficients.

\section{Pseudospectral method}
\label{sec:pseudospectral}

In this section, we review how one starts with the ODE eigenvalue problem in \eqref{eq:fullproblem} and end up with the generalized eigenvalue problem in \eqref{eq:GenericMatrixEquation}. We derive a general recipe for mapping a differential operator and function pair to a matrix and vector pair $(\tilde{\bm{\mathcal{M}}}(\omega),\tilde{\bm{\mathcal{C}}})$ via a collocation method in the Bernstein basis, whose closed form can be found in the last section. In the context of Chebyschev basis polynomials and Fourier basis functions, the standard reference is \cite{BoydSpectral}.

We start with a linear eigenvalue ODE, then show how it can be extended to polynomial eigenvalue ODEs. We extend this to include problems involving a set of dependent functions. We elaborate on special cases in \ref{sec:appendix}, used to convert the polynomial generalized eigenvalue problem to an eigenvalue problem.

\subsection{Linear eigenvalue problem}
\label{subsect:LEP}

Consider the ODE eigenvalue problem in \eqref{eq:probsol}, specifically of the form
\begin{equation}	
\hat{L}(u,\omega) \psi(u) = (\hat{f}_0(u) + \omega \hat{f}_1(u))\psi(u) = 0.
\label{eq:linearODE}
\end{equation}
To arrive at a spectral matrix of size $N+1$, we expand the basis degree to $N_{max} = N + q + r$.
\begin{equation}
\psi(u) \approx \sum_{k = 0}^{N} C_{k+q} B^{N_{\mathrm{max}}}_{k+q}(u).
\label{eq:linearODEsol}
\end{equation}
\begin{figure}[!t]
\begin{centering}
\begin{tabular}{c}
\includegraphics[width=0.45\textwidth]{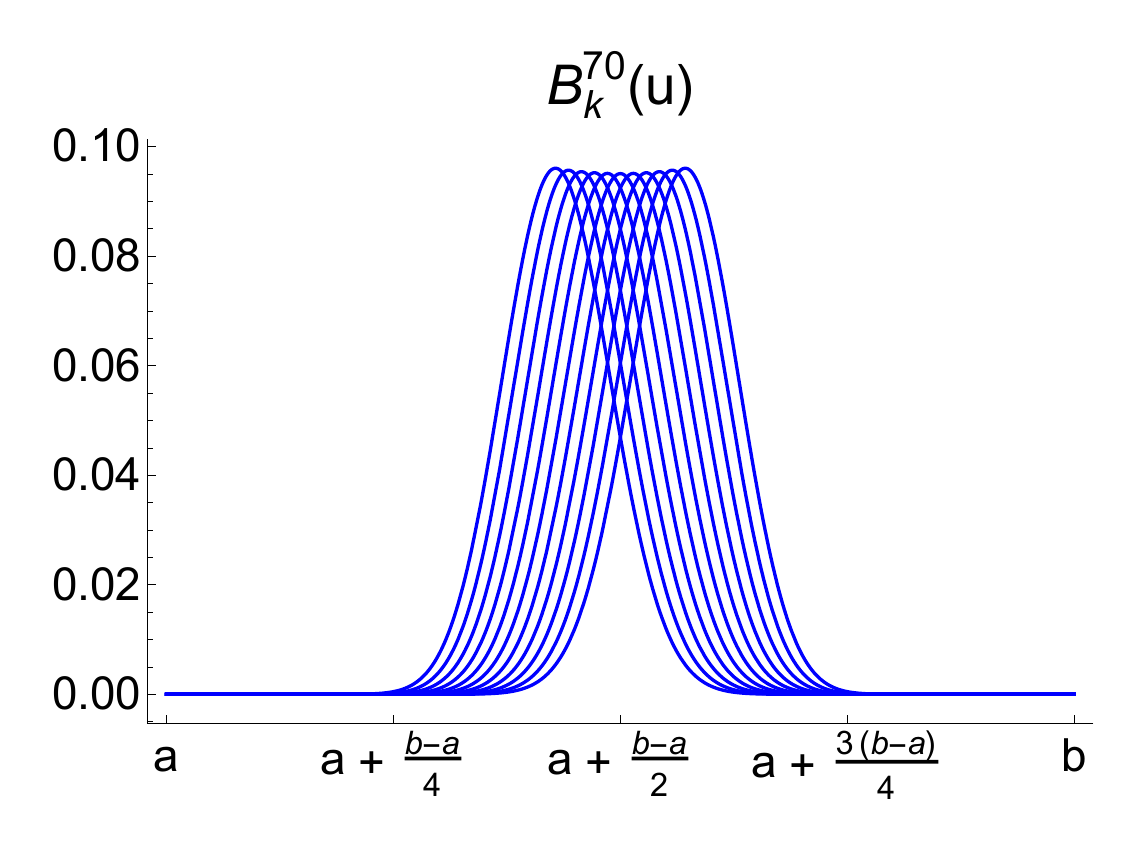} \\
\includegraphics[width=0.45\textwidth]{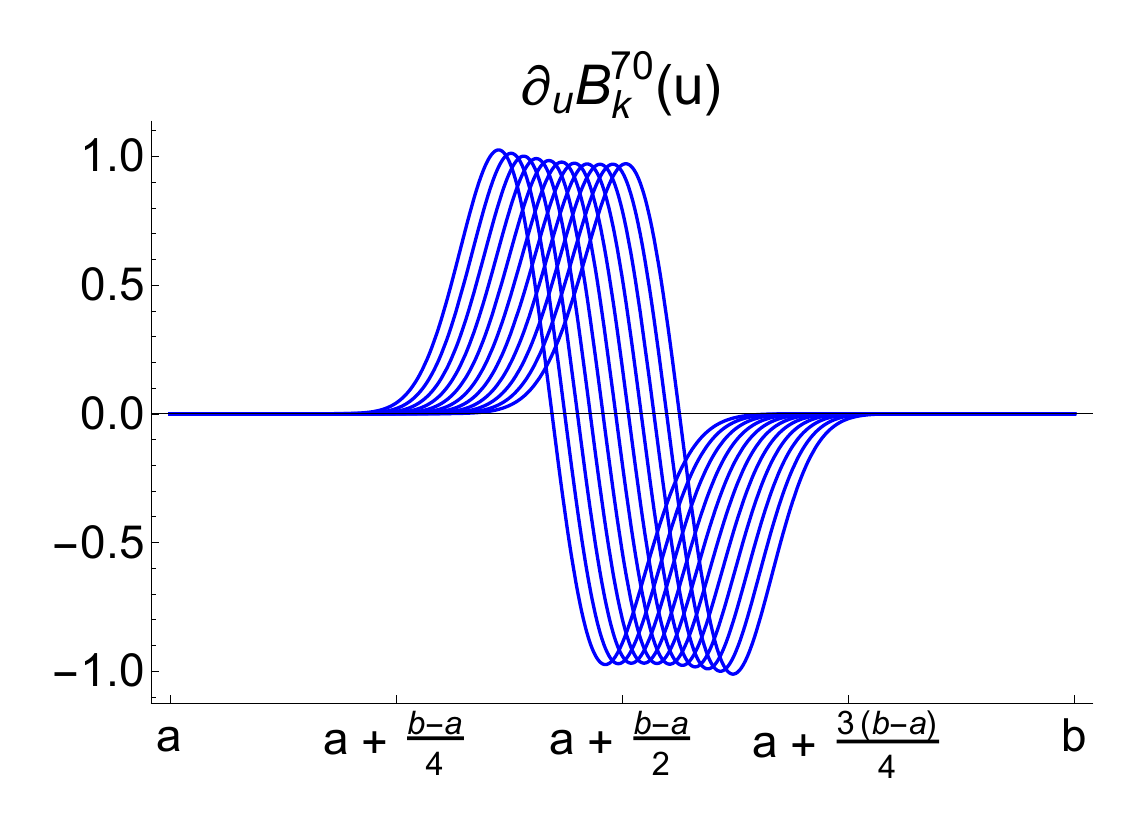} \\
\includegraphics[width=0.45\textwidth]{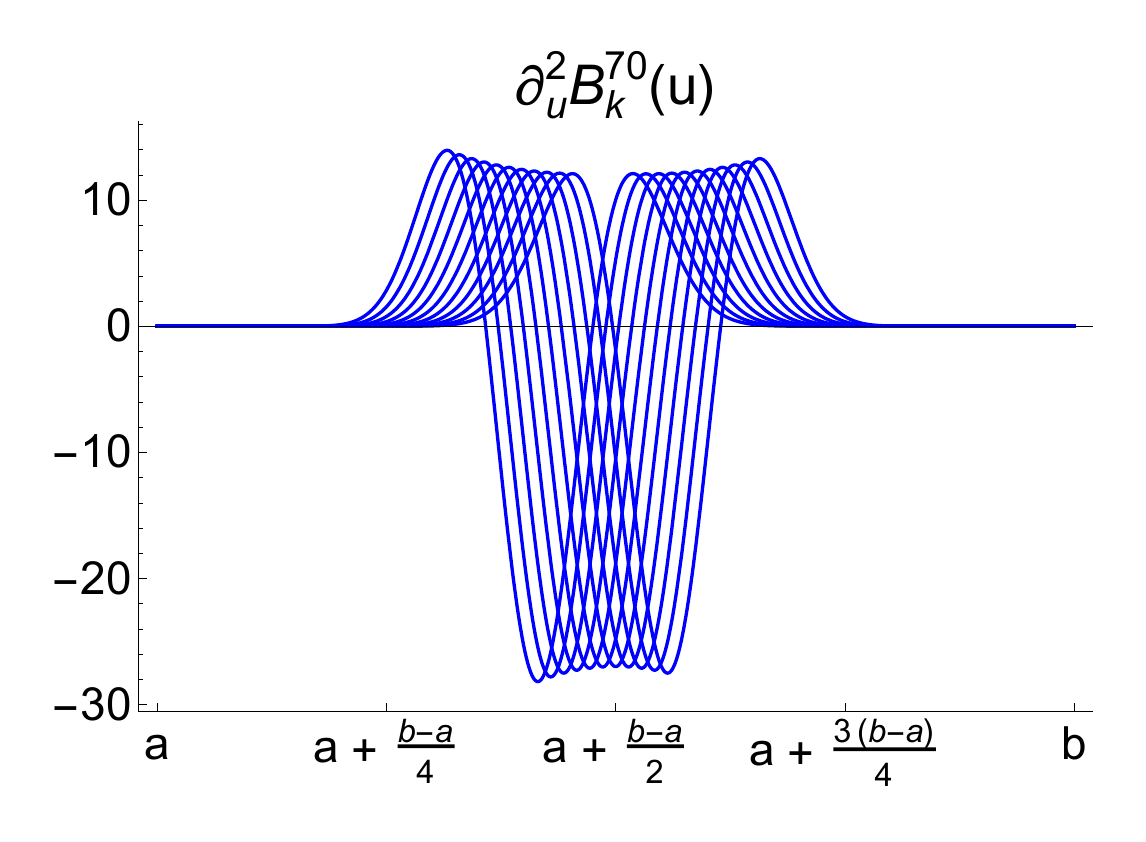} \\
\end{tabular}
\caption{The set of 11 Bernstein basis polynomials appropriate when $q = 30$ and $r = 30$, and their derivatives. The basis functions are localized around the center of $[a,b]$, as are their derivatives.}
\label{fig:BP301030}
\end{centering}
\end{figure}
A straightforward implementation of the collocation method would be to define a grid of $N+1$ points in the interval $[a,b]$. Since the first $q$ or last $r$ Bernstein basis functions dominate the behaviour of the solution near the boundaries, we propose instead to select collocating points in the region dominated by the basis functions whose weights are still unknown.

As an illustrative example, consider the case when $N = 10$, $q = 30$ and $r = 30$. One can imagine rescaling a solution $\phi(u)$ finite at both boundaries via the transformation,
\begin{equation}
\phi(u) = \dfrac{\tilde{\phi}(u)}{(u-a)^{30}(b-u)^{30}}.
\label{eq:rescaling}
\end{equation}
The basis of $\phi(u)$ is in Figure \ref{fig:BP10} while the basis of $\tilde{\phi}(u)$ is in Figure \ref{fig:BP301030}. Its derivatives are similarly localized.
We construct our collocating grid by considering a Chebyschev or equally spaced grid of $N_{\mathrm{max}} + 1$ points over $[a,b]$, 
\begin{equation}
\{u_0, u_1, \dots, u_{N_{\mathrm{max}}}\}, \qquad u_0 = a, \; \; \;u_{N_{\mathrm{max}}}=b,
\end{equation}
and then select grid points $q$ through $N + q + 1$. 

Let us now endeavour to convert the differential operator and function pair $(\hat{f}(u),\psi(u))$ into a matrix and vector pair $(\bm{M},\bm{C})$. Suppose $\hat{f}(u)$ is of the form,
\begin{equation}
\hat{f}(u) = \sum_{n=0}^{n_{\mathrm{max}}} f_n(u) \deri{^n}{u^n}.
\end{equation}
A generic term in the $\hat{f}(u)\psi(u)$ is of the form $f_n(u) \deri{^n \psi}{u^n}$. Combining \eqref{eq:BPderivative} and \eqref{eq:linearODEsol}, we get 
\begin{multline}
f_n(u) \deri{^n \psi(u)}{u^n} = \dfrac{f(u)}{(b-a)^n} \dfrac{(N_{\mathrm{max}})!}{(N_{\mathrm{max}}-n)!} \times \\
\sum_{k = 0}^{N} \sum_{l=0}^n (-1)^l \take{n}{l} B^{N_{\mathrm{max}}-n}_{k+q+l-n}(u) C_{k+q}.
\label{eq:linearODEcoll1}
\end{multline}
%\begin{equation}
%f_n(u) \deri{^n \psi(u)}{u^n} = \dfrac{f(u)}{(b-a)^n} \dfrac{(N_{\mathrm{max}})!}{(N_{\mathrm{max}}-n)!}
%\sum_{k = 0}^{N} \sum_{l=0}^n (-1)^l \take{n}{l} B^{N_{\mathrm{max}}-n}_{k+q+l-n}(u) C_{k+q}.
%\label{eq:linearODEcoll1}
%\end{equation}
We may assign a vector to each term in $\hat{f}(u)\psi(u)$ with the condition that the differential operator is satisfied at each collocation point,
\begin{equation}
f_n(u) \deri{^n \psi(u)}{u^n} \to \bm{T}^{(n)} \bm{C},
\end{equation}
where $\bm{C}_k = C_{k+q}$ and the matrix components of $\bm{T}^{(n)}$ are given by
\begin{multline}
\bm{T}^{(n)}_{j,k} = \dfrac{f_n(u_{j+q})}{(b-a)^n} \dfrac{(N_{\textnormal{max}})!}{(N_{\textnormal{max}}-n)!} \times \\
\sum_{l=0}^n (-1)^l \take{n}{l} B^{N_{\textnormal{max}}-n}_{k+q+l-n}(u_{j+q}),
\label{eq:preliminaryspectralmatrix}
\end{multline}
%\begin{eqnarray}
%\bm{T}^{(n)}_{j,k} = \dfrac{f_n(u_{j+q})}{(b-a)^n} \dfrac{(N_{\mathrm{max}})!}{(N_{\mathrm{max}}-n)!}
%\sum_{l=0}^n (-1)^l \take{n}{l} B^{N_{\mathrm{max}}-n}_{k+q+l-n}(u_{j+q}),
%\label{eq:preliminaryspectralmatrix}
%\end{eqnarray}
for $j,k \in \{0,1,\dots,N\}$.

To use \eqref{eq:preliminaryspectralmatrix}, each Bernstein basis polynomial of degree \linebreak $N_{\mathrm{max}}-n$ through $N_{\mathrm{max}}$ must be evaluated at each collocation point. Since in many applications, $n_{\mathrm{max}} \ll N$, it would be numerically cost efficient to use \eqref{eq:BPraising} and rewrite \eqref{eq:preliminaryspectralmatrix} in terms of a single Bernstein basis degree, as in
\begin{multline}
\bm{T}^{(n)}_{j,k} = \dfrac{f_n(u_{j+q})}{(b-a)^n} \dfrac{(N_{\textnormal{max}})!}{(N_{\textnormal{max}}-n)!}\sum_{l=0}^n (-1)^l \take{n}{l} \times\\
 \sum_{m=0}^n \dfrac{\take{n}{m} \take{N_{\textnormal{max}}-n}{k+q+l-n}}{\take{N_{\textnormal{max}}}{k+q+l+m-n}} B^{N_{\textnormal{max}}}_{k+q+l+m-n}(u_{j+q}).
 \label{eq:spectralmatrix1}
\end{multline}
%\begin{equation}
%\begin{array}{ll}
%\bm{T}^{(n)}_{j,k} = \dfrac{f_n(u_{j+q})}{(b-a)^n} & \dfrac{(N_{\mathrm{max}})!}{(N_{\mathrm{max}}-n)!}\sum_{l=0}^n (-1)^l \take{n}{l} \times \nonumber \\
%& \sum_{m=0}^n \dfrac{\take{n}{m} \take{N_{\mathrm{max}}-n}{k+q+l-n}}{\take{N_{\mathrm{max}}}{k+q+l+m-n}} B^{N_{max}}_{k+q+l+m-n}(u_{j+q}).
%\end{array}
% \label{eq:spectralmatrix1}
%\end{equation}
By choosing this degree to be $N_{\mathrm{max}}$, only a subset of the Bernstein basis needs to be evaluated at each collocation point$-$specifically those indexed in the range \linebreak $[q - \min(n_{\mathrm{max}},q), N + q + \min(n_{\mathrm{max}},r)]$. 
Thus,
\begin{equation}
(\hat{f}(u),\psi(u)) \to (\bm{M},\bm{C}), \qquad \bm{M} = \sum_{n=0}^{n_{\mathrm{max}}} \bm{T}^{(n)}.
\end{equation}
The ODE linear eigenvalue problem in \eqref{eq:linearODE} may be written as a generalized eigenvalue problem,
\begin{equation}
\bm{M}(\omega) \bm{C} = (\bm{M}_0 + \omega \bm{M}_1)\bm{C} = 0.
\end{equation}

\subsection{Polynomial eigenvalue problem}
\label{subsect:PEP}
Consider a polynomial eigenvalue problem of order $m$.,
\begin{equation}
 (\hat{f}_0(u) + \omega \hat{f}_1(u) + \omega^2 \hat{f}_2(u) + \dots + \omega^m \hat{f}_m(u) )\psi(u) = 0.
\end{equation}
Using the recipe discussed in the previous section, this corresponds to an eigenvalue problem of a \textit{matrix pencil} of order $m$,
\begin{equation}
(\bm{M}_0 + \omega \bm{M}_1 + \omega^2 \bm{M}_2 + \dots + \omega^m \bm{M}_m)\bm{C} = 0.
\label{eq:polynomialmatrix}
\end{equation}
We linearize the matrix pencil by defining the following matrices,
\begin{equation}
\bm{\mathcal{M}}' = \left(\begin{array}{cccc}
\bm{M}_0 & \bm{M}_1 & \dots & \bm{M}_{m-1} \\
0 & \mathds{1} & \dots & 0 \\
\vdots & \vdots & \ddots & \vdots \\
0 & 0 & \dots & \mathds{1} \\
\end{array}\right),
\end{equation}
\begin{equation}
\bm{\mathcal{M}}'' = \left( \begin{array}{cccc}
0 & \dots & 0 & \bm{M}_m \\
-\mathds{1} & \dots & 0 & 0 \\
\vdots & \ddots & \vdots & \vdots \\
0 & \dots & -\mathds{1} & 0 \\
\end{array} \right),
\end{equation}
\noindent and the vector,
\begin{equation}
\bm{\mathcal{C}} = \left( \begin{array}{c}
\bm{C} \\
\omega \bm{C} \\
\vdots \\
\omega^{m-1} \bm{C} \\
\end{array} \right). 
\end{equation}
This transforms the matrix pencil \eqref{eq:polynomialmatrix} to another GEP,
\begin{equation}
\bm{\mathcal{M}}(\omega) \bm{\mathcal{C}} = (\bm{\mathcal{M}}' + \omega \bm{\mathcal{M}}'') \bm{\mathcal{C}} = 0.
\label{eq:polyGEP}
\end{equation}
For clarity, we typeset matrices and vectors generated from linearizing a matrix pencil by a calligraphic typeface. 

Generalized eigenvalue problems are more difficult to solve than regular eigenvalue problems. We describe a method to transform the above GEP to an EP in \ref{subsec:GEPtoEP} contingent on the invertibility of either $\bm{M}_0$ or $\bm{M}_m$, leading to a modest improvement in speed.

\subsection{Polynomial eigenvalue problem over several dependent functions}
\label{subsect:fullPEP}

Consider the full problem in Section \ref{sec:introduction}. In matrix form, this becomes the set of simultaneous equations,
\begin{eqnarray}
\bm{\mathcal{M}}_{1,1}(\omega) \bm{\mathcal{C}}_1 + \dots + \bm{\mathcal{M}}_{1,n}(\omega) \bm{\mathcal{C}}_n &=& 0, \nonumber \\
\bm{\mathcal{M}}_{2,1}(\omega) \bm{\mathcal{C}}_1 + \dots + \bm{\mathcal{M}}_{2,n}(\omega) \bm{\mathcal{C}}_n &=& 0, \nonumber \\
&\vdots & \label{eq:fullPEP1}\\
\bm{\mathcal{M}}_{n,1}(\omega) \bm{\mathcal{C}}_1 + \dots + \bm{\mathcal{M}}_{n,n}(\omega) \bm{\mathcal{C}}_n &=& 0, \nonumber
\end{eqnarray}
where each matrix $\bm{\mathcal{M}}_{j,k}(\omega)$ is constructed by linearizing the matrix pencil of the $k$th dependent function of the $j$th equation, as in
\begin{equation}
\bm{\mathcal{M}}_{j,k}(\omega) = \bm{\mathcal{M}}_{j,k}' + \omega \bm{\mathcal{M}}_{j,k}''.
\end{equation}
The set of simultaneous equations can be written as a single matrix equation by defining the following  matrices,
\begin{equation}
\tilde{\bm{\mathcal{M}}}' = \left( \begin{array}{cccc}
\bm{\mathcal{M}}'_{1,1} & \bm{\mathcal{M}}'_{1,2} & \dots & \bm{\mathcal{M}}'_{1,n} \\
\bm{\mathcal{M}}'_{2,1} & \bm{\mathcal{M}}'_{2,2} & \dots & \bm{\mathcal{M}}'_{2,n} \\
\vdots & \vdots & \ddots & \vdots \\
\bm{\mathcal{M}}'_{n,1} & \bm{\mathcal{M}}'_{n,2} & \dots & \bm{\mathcal{M}}'_{n,n} \\
\end{array} \right),
\end{equation}
\begin{equation}
\tilde{\bm{\mathcal{M}}}'' = \left( \begin{array}{cccc}
\bm{\mathcal{M}}''_{1,1} & \bm{\mathcal{M}}''_{1,2} & \dots & \bm{\mathcal{M}}''_{1,n} \\
\bm{\mathcal{M}}''_{2,1} & \bm{\mathcal{M}}''_{2,2} & \dots & \bm{\mathcal{M}}''_{2,n} \\
\vdots & \vdots & \ddots & \vdots \\
\bm{\mathcal{M}}''_{n,1} & \bm{\mathcal{M}}''_{n,2} & \dots & \bm{\mathcal{M}}''_{n,n} \\
\end{array} \right),
\end{equation}
and vector,
\begin{equation}
\tilde{\bm{\mathcal{C}}} = \left( \begin{array}{c}
\bm{\mathcal{C}}_1 \\
\bm{\mathcal{C}}_2 \\
\vdots \\
\bm{\mathcal{C}}_n
\end{array} \right).
\end{equation}
We arrive at the GEP of the full problem introduced in Section \ref{sec:introduction},
\begin{equation}
\tilde{\bm{\mathcal{M}}}(\omega)\tilde{\bm{\mathcal{C}}} = (\tilde{\bm{\mathcal{M}}}' + \omega \tilde{\bm{\mathcal{M}}}'')\tilde{\bm{\mathcal{C}}} = 0.
\label{eq:fullmatrixproblem2}
\end{equation}
The GEP of the full problem is much more complicated. Unlike the GEP of the previous subsection, it can be shown that $\tilde{\bm{\mathcal{M}}}'$ is always singular, as we show in \ref{subsec:GEPtoEP}.

\section{Advantages and disadvantages of the Bernstein basis}
\label{sec:procon}

Having elaborated on how the Bernstein basis fits into solving a partial differential problem like \eqref{eq:fullproblem}, we discuss in this section what these properties cost and afford us, and how they compare to more standard basis functions. We also discuss some results that may be found in \ref{subsec:eigenman}. One may read through that Section first, and then return here.
\begin{enumerate}
\item Bernstein polynomials are not orthogonal. This follows from \eqref{eq:BPint} and \eqref{eq:BPproduct}. This complicates an extension of the current method to partial differential equations, where the weights may be made to vary in time.
\item The Bernstein basis polynomials depends on the basis degree. We cannot naively apply derivatives without costing us additional numerical resources. We need to fold in an application of \eqref{eq:BPraising} so that we remain in a single common basis degree. There is no operation similar to \eqref{eq:BPraising} for classical orthogonal polynomials, because those basis functions do not depend on the size of the basis.
\item The zeros of the Bernstein basis, if they occur, are located at the boundaries. There are no nodes we can take advantage of in constructing a collocation grid, so the implemented spectral matrices \eqref{eq:implementedspecmatrix1} and \eqref{eq:implementedspecmatrix2} are dense.
\item Many of the properties of the Bernstein basis have equivalent forms for other basis functions. The boundary values, derivative recurrence relation and integral similar to \eqref{eq:boundary}, \eqref{eq:BPderivative} and \eqref{eq:BPint} are well-known for classical orthogonal polynomials and the Fourier basis. A simple product formula like \eqref{eq:BPproduct} exists for Chebyschev and Fourier basis.
\item The specific form of these properties gives the Bernstein basis an advantage over other basis functions when dealing with mixed boundary value problems outlined in Section~\ref{sec:boundary}. In the Bernstein basis, the boundary constraints only act on a subset of the basis set, whose weights can be fully determined independently of the differential equation. Such a luxury is not enjoyed by more standard basis functions. For classical orthogonal polynomials and the Fourier basis, imposing the boundary conditions \eqref{eq:BoundaryConditions1} would involve the entire basis set. Solving the differential equations and the boundary conditions must be done simultaneously.
\item The lack of a residual term in \eqref{eq:probsol} and the lack of additional constraints on the expansion coefficients lets us write down the algebraic equations these expansion coefficients must satisfy as a generalized eigenvalue problem in Section~\ref{sec:pseudospectral}.
\item There are manipulations which can only be easily done in the Bernstein basis, discussed in \ref{subsec:eigenman}. For example, a tau method using Chebyschev polynomials can impose the boundary condition $\lim_{u \to a} \psi(u) \sim (u-a)$ exactly. However, one cannot naively divide out a $(u-a)$ term-by-term, since each Chebyschev polynomial is finite at the lower boundary. Such a rescaling can be exactly carried out in the Bernstein basis, as shown in \ref{subsec:eigenman}. This lets us calculate the weighted $L^2$-norm of a function in the Bernstein basis in closed-form, even in cases where the weight has a pole of integer degree at the boundaries. This is useful, for example, when normalizing wavefunctions in a compactified coordinate system, as in Section~\ref{subsec:QHO}.
\item The numerical convergence of the Bernstein basis has been established in the context of other differential problems. Interestingly, in some cases, the Bernstein method would outperform other basis functions (including Chebyshev and Fourier) in terms of numerical cost or the accuracy of the solutions \cite{Yuzbasi2013,Maleknejad2012,Maleknejad2012a,Hesameddini2017,Yuzbasi2016}. We do not perform a similarly comprehensive analysis here, concentrating instead on general ideas on how the Bernstein basis may be adapted to ODE eigenvalue problems and introducing the package \texttt{SpectralBP}. Though we do demonstrate numerical convergence for some of the cases we tackle below.

\end{enumerate}

\section{The \texttt{SpectralBP} package}
\label{sec:SBPintro}

The \texttt{SpectralBP} package uses the properties of the Bernstein basis, written to streamline the calculation of the eigenvalues and eigenfunctions of \eqref{eq:fullproblem}. It is primarily distributed as a \texttt{Mathematica} paclet and is publicly available \cite{spectralbp}.

\texttt{SpectralBP} commands are documented, and the package is bundled with two tutorial notebooks. After installation, the details and options of each command may be explored by prefixing a command with a question mark, as in \texttt{?GetModes}, similar to built-in commands in \texttt{Mathematica}.

There are three types of commands in \texttt{SpectralBP}: \texttt{Get} commands, \texttt{Compare} commands and \texttt{Print} commands. The basic work flow is as follows.
\begin{enumerate}
\item Begin with some ODE eigenvalue problem
\begin{equation}
\hat{L}'(x,\omega)\Psi'(x) = 0
 \end{equation} 
which may not satisfy the 3 properties required in Section \ref{sec:introduction}.
\item If the domain of the eigenfunctions $\psi'_i(x)$ is not compact, define an invertible change of variables $f(x) = u$ so that the domain in $u$ is compact.
\item If the resulting eigenfunctions are non-analytic, one may rescale as in
\begin{equation}
 \psi'_i(u) = f_i(u) \psi_i(u)
 \end{equation}
so that the resulting eigenfunctions $\psi_i(u)$ are analytic. 

One also defines $f_i(u)$ so that all eigenfunctions $\psi_i(u)$ satisfies the same boundary conditions. The result should be an eigenvalue problem described in Section \ref{sec:introduction}.
\item Use \texttt{Get} commands to calculate eigenvalues and eigenfunctions at different BP orders.
\item Use \texttt{Compare} commands to filter out spurious eigenvalues and eigenfunctions.
\item Use \texttt{Print} commands to quickly glean off information from the prior calculations.
\end{enumerate}

We will discuss each command type in the following subsection before going into applications. Example notebooks can be found in the next two Sections. 

\subsection{\texttt{Get} commands}
\label{subsec:getcommands}

The first input of a \texttt{Get} command is a list of differential equations. The command automatically identifies the dependent functions, the independent variable and the eigenvariable. The command halts whenever it identifies more than one independent variable or eigenvariable, or whenever the number of dependent functions underdetermine or overdetermine the problem.

There are three \texttt{Get} commands,
\begin{enumerate}
\item \texttt{GetModes[eqn,N]}: Calculates the eigenvalues of the ODE eigenvalue problem stored in \texttt{eqn} using a basis degree of \texttt{N}. 
\item \texttt{GetEigenfunctions[eqn,modes,N]}: Calculates the eigenvectors corresponding to each eigenvalue in the list \texttt{modes}, using a basis degree of \texttt{N}. As discussed in the Appendix, we advise that \texttt{N} be identical to be basis degree the eigenvalues in \texttt{modes} were calculated.
\item \texttt{GetAccurateModes[eqn,N1,N2]}: Calculates the eigenvalues using basis degrees of \texttt{N1} and \texttt{N2}, then applies a \texttt{CompareModes} command to filter the eigenvalues.
\end{enumerate}

By replacing the basis degree inputs with a pair of numbers, which we call a \textit{basis tuple} of the form \texttt{\{N,prec\}}, eigenvalues are calculated using a basis degree of \texttt{N} using \texttt{prec}-precision numbers. That is, an alternative input scheme for the above commands is given by,
\begin{eqnarray*}
&\texttt{GetModes[eqn,\{N,prec\}]}, \\
&\texttt{GetAccurateModes[eqn,\{N1,prec1\},\{N2,prec2\}]}.
\end{eqnarray*}

The default behavior of $\texttt{GetModes}$ and $\texttt{GetAccurateModes}$ are as follows
\begin{equation*}
\texttt{GetModes[eqn,N]} = \texttt{GetModes[eqn,\{N,N/2\}]}, 
\end{equation*}
\begin{multline*}
\texttt{GetAccurateModes[eqn,N1,N2]}  = \\
 \texttt{GetAccurateModes[eqn,\{N1,N1/2\},\{N2,N2/2\}]}.
\end{multline*}

In calculating the eigenvalues and eigenvectors, \texttt{Get} commands must be supplied with the correct domain and boundary conditions. These are controlled by 4 options,
\begin{enumerate}
\item \texttt{LowerBound} and \texttt{UpperBound}: defines the domain $[a,b]$, which defaults to $[0,1]$.
\item \texttt{LBPower} and \texttt{UBPower}: defines the leading polynomial power $q,r$ at each boundary, which defaults to $q = 0$ and $r = 0$.
\end{enumerate}

The option \texttt{Normalization} lets one choose how eigenfunctions are normalized. The option may have 4 values,
\begin{enumerate}
\item \texttt{"UB"}: the coefficient of the leading polynomial expansion of the eigenfunctions at $b$ to 1.
\item \texttt{"LB"}: the coefficient of the leading polynomial expansion of the eigenfunctions at $a$ to 1.
\item \texttt{"L2Norm"}: the $L^2$-norm of the eigenfunctions to 1.
\item \texttt{\{"L2Norm",\{A,B,C\}\}}: the $L^2$-norm of the eigenfunctions to 1, with a weight function underneath the integral of the form $A (u-a)^B (b-u)^C$.
\end{enumerate}

The option \texttt{FinalAsymptotics} lets one change the outputted eigenfunctions' asymptotics, according to manipulations detailed in \ref{subsec:eigenman}.

\begin{table*}[!t]
\caption{\label{table:summary} Input scheme for the various eigenvalue ODE problems discussed in Section~\ref{sec:TISE}, Section~\ref{sec:QNM} and Section~\ref{sec:ASM}. The potential $V^*$ was chosen to be \eqref{eq:ISWpotential} for the base infinite square well problem, and \eqref{eq:ISWpertpot} for the perturbed infinite square well problem. The potential $V^\dagger$ was chosen to be \eqref{eq:QHOpotential} for the quantum harmonic oscillator problem. The potential $V^\ddagger$ was chosen to be \eqref{eq:anharmonic1} as the $\mathcal{PT}$-symmetric anharmonic potential for specific values of $\lambda$, and \eqref{eq:anharmonic2} as the quartic anharmonic potential for specific values of $\beta$. The different variables used mark certain coordinate transformations effected to compactify an infinite domain.}
{\footnotesize
\begin{center}
\begin{tabular}{llcc}
\hline \hline
Problem & \texttt{eqn} & \texttt{LBPower} & \texttt{UBPower}\\
 \hline \hline 
Infinite square well & $\dfrac{1}{2} \phi''(x) + (E - V^*(x)) \phi(x)$ & 1 & 1 \vspace{3pt}\\
\multirow{2}{*}{Harmonic oscillator} & $ \dfrac{1}{2}v_2^2(v_2-1)^2\phi''(v_2) + \dfrac{1}{2}v_2(v_2-1)(2v_2 - 1) \phi'(v_2)~+$ & \multirow{2}{*}{1} & \multirow{2}{*}{1} \vspace{3pt}\\
 & $\qquad \qquad \left( E - V^\dagger\left(\ln \left[ \dfrac{v_2}{1-v_2}\right] \right) \right) \phi(v_2)$ &  & \vspace{3pt}\\
\multirow{2}{*}{Anharmonic oscillator} & $ v_2^2 (v_2-1)^2 \phi''(v_2) + v_2 (v_2-1)(2v_2 - 1) \phi'(v_2)~+$ & \multirow{2}{*}{1} & \multirow{2}{*}{1} \vspace{3pt}\\
 & $\qquad \qquad  \left( E - V^\ddagger\left(\ln \left[ \dfrac{v_2}{1-v_2}\right] \right) \right) \phi(v_2)$ &  & \vspace{3pt}\\
 \multirow{2}{*}{Schwarzschild QNMs} & $ (1-u) u^2 \phi''(u) + (2 \lambda + 2 u - u^2 (3 + 4 \lambda)	 )\phi'(u)~-$ & \multirow{2}{*}{0} & \multirow{2}{*}{0} \vspace{3pt}\\
 & $\qquad \qquad  \left(l + l^2 + 4 \lambda^2 + u ((1 + 2 \lambda)^2 - s^2)\right) \phi(u)$ &  & \\
 \hline
\end{tabular}
\end{center}
}
\end{table*}

\subsection{\texttt{Compare} commands}
\label{subsec:comparecommands}

The spectrum calculated from a finite basis degree will be filled with either eigenvalues that have not converged or spurious eigenvalues. We have provided two ways to filter these out. These are the two \texttt{Compare} commands,
\begin{enumerate}
\item \texttt{CompareModes[modes1,modes2]}: Checks whether eigenvalues in the two spectra inputted share common digits, then keeps only eigenvalues that share at least $3$ digits.
\item \texttt{CompareEigenfunctions[eqn,\{modes1,\linebreak modes2\},\{N1,N2\}]}: Calculates the eigenfunctions of the eigenvalues approximately common to \texttt{modes1} and \texttt{modes2} using a basis degree of \texttt{N1} and \texttt{N2} respectively. If the $L^2$-norm of their difference is less than $10^{-3}$, the eigenvalues are kept.
\end{enumerate}

There are two relevant options,
\begin{enumerate}
\item \texttt{Cutoff}: controls the minimum number of common digits for eigenvalues to be kept, which defaults to 3.
\item \texttt{L2Cutoff}: controls the maximum difference between two eigenfunctions, of the form $10^{-n}$, for their corresponding eigenvalues to be kept, which defaults to $n=3$.
\end{enumerate}

We call eigenvalues of different spectra that share a \texttt{Cutoff}-number of common digits \textit{approximately common}.

One may also input a list of spectra into \texttt{CompareModes}, as in
\begin{equation*}
\texttt{CompareModes[\{modes1,modes2,\dots\}]}.
\end{equation*}

\subsection{\texttt{Print} commands}
\label{subsec:printcommands}

There are four \texttt{Print} commands,
\begin{enumerate}
\item \texttt{PrintFrequencies[modes]}: plots the eigenvalues in \texttt{modes} on the complex plane.
\item \texttt{PrintEigenfunctions[eqn,modes,N]}: plots the real and imaginary parts of the corresponding eigenfunctions.
\item \texttt{PrintTable[convergedmodes]}: generates a table of eigenvalues, categorizing them into purely real, purely imaginary, and complex eigenvalues. Groups together eigenvalues satisfying $\omega^* = \omega$ and $\omega^* = - \omega$. The input must be a pair of lists of approximately common eigenvalues, usually coming from the output of a \texttt{CompareModes} command.
\item \texttt{PrintAll[eqn,convergedmodes,N]}: a shortcut to do the previous three commands in a single command.
\end{enumerate}

There are three relevant options,
\begin{enumerate}
\item \texttt{FreqName}: specifies the symbol for the eigenvariable, which defaults to $\omega$.
\item \texttt{NSpectrum}: specifies how many eigenvalues would be plotted, which defaults to plotting everything.
\item \texttt{NEigenFunc}: specifies how many eigenfunctions would be plotted, which defaults to plotting everything.
\end{enumerate}

The \texttt{PrintTable} command automatically only prints out \textit{significant~digits}, defined to be the digits common to both the spectra inputted. When the inputted spectra comes from two adjacent basis degrees, say $N$ and $N+1$, the right-most digits of the output may be incorrect. This occurs because the absolute error of the two spectra overlap.

We recommend using basis degrees that are far apart in the sense that the absolute error of the higher basis degree spectrum is much smaller than the absolute error of the lower basis degree spectrum. Although the practice would be numerically more costly, in this way we increase our chances that the right-most significant digit outputted is correct.

\subsection{Summary of implementations}

\floatname{algorithm}{Notebook}
\begin{table*}
\begin{minipage}{\textwidth}
\begin{algorithm}[H]
\caption{: A simple \texttt{Mathematica} notebook implementation of \texttt{SpectralBP} for the infinite square well problem.}\label{alg:basiccommands}
\begin{algorithmic}[1]
\State TISE =  Equation \eqref{eq:TISE} with potential \eqref{eq:ISWpotential}
\State modes50 = GetModes[TISE, 50, LBPower$\to$1, UBPower$\to$1] \label{algline:getmodes1}
\State modes80 = GetModes[TISE, 80, LBPower$\to$1, UBPower$\to$1] \label{algline:getmodes2}
\State convergedmodes = CompareModes[modes50, modes80]  \label{algline:comparemodes}
\State PrintFrequencies[$\dfrac{2}{\pi^2}$modes50, NSpectrum$\to$10, FreqName$\to$`$\dfrac{2}{\pi^2}$E'] \Comment{Figure \ref{fig:quickmodes} (Top)}\label{algline:printfrequencies}
\State PrintEigenfunctions[TISE, modes[[1;;3]], 50, LBPower$\to$1, UBPower$\to$1, Normalization$\to$`L2Norm'] \Comment{Figure \ref{fig:quickmodes} (Middle)}\label{algline:eigenfunctions}
\State PrintTable[$\dfrac{2}{\pi^2}$convergedmodes[[;;,1;;10]], FreqName$\to$`$\dfrac{2}{\pi^2}$E'] \Comment{Figure \ref{fig:quickmodes} (Bottom)} \label{algline:printtable}
\end{algorithmic}
\end{algorithm}
\end{minipage}
\end{table*}

In Table~\ref{table:summary}, we summarize the different inputs needed to solve the ODE eigenvalue problems that we shall look at in the succeeding Sections. Hopefully, in the examples considered in the proceeding Sections, one is left with an impression of the general-purpose applicability and ease-of-use of \texttt{SpectralBP}. As shall be demonstrated, three lines of code can yield a wealth of information about the considered ODE eigenvalue problem. The difference between the examples given is just swapping in and out of differential equations, applying certain change of variables in cases where the domain is infinite, and specifying the necessary boundary conditions.

\section{Applications in Quantum Mechanics}
\label{sec:TISE}

We first illustrate how \texttt{SpectralBP} is used by working through standard problems in quantum mechanics. We solve for the eigenenergies and eigenfunctions of the infinite square well and quantum harmonic potentials numerically in the first two subsection. Calculations are compared with well-known analytic results, as can be found in standard quantum mechanics textbooks like \cite{GriffithsQM}.

For the last two subsections, we compute the eigenenergies of the anharmonic potentials considered in \cite{Bender2001a} and \cite{Ezawa2014}. We compare ground state eigenenergies calculated with \texttt{SpectralBP} to the results of the aforementioned papers, which were both calculated perturbatively using a combination of Pad\'{e} approximation and Stieltjes series. In \cite{Ezawa2014}, Milne's method \cite{Milne1930} was used as an independent test.

\subsection{Infinite square well}
Consider the time-independent Schr\"odinger equation
\begin{equation}
\dfrac{1}{2}\deri{^2}{x^2}\phi(x) + (E - V(x))\phi(x) = 0.
\label{eq:TISE}
\end{equation}
For the infinite square well, the potential is chosen to be
\begin{equation}
V(x) = \left\lbrace 
\begin{array}{ll}
0 & 0 \leq x \leq 1 \\
\infty & \mathrm{otherwise}.
\end{array}
\right.
\label{eq:ISWpotential}
\end{equation}
Its eigenenergies are
\begin{equation}
E_n = \dfrac{\pi^2 n^2}{2}, \qquad n = 1,2,3,\dots.
\label{eq:ISWenergies}
\end{equation}
The domain of solutions is the interval $[0,1]$ with boundary conditions,
\begin{equation}
\lim_{x \to 0} \phi(x) \sim x, \qquad \lim_{x \to 1} \phi(x) \sim (1-x).
\label{eq:ISWbounds}
\end{equation}

\subsubsection{\texttt{SpectralBP}$-$basic implementation.}

A simple implementation to solve the infinite square well problem is schematically found in Notebook \ref{alg:basiccommands}.

Lines \ref{algline:getmodes1} and \ref{algline:getmodes2} solves the ODE eigenvalue problem \eqref{eq:TISE} with potential \eqref{eq:ISWpotential} using basis degrees 50 and 80 respectively. 

The boundary conditions \eqref{eq:ISWbounds} are set by the option values
\begin{equation*}
\texttt{LBPower $\to$ 1}, \qquad \texttt{UBPower $\to$ 1},
\end{equation*}
\noindent which must be specified whenever eigenvalues and eigenvectors are calculated.

Line \ref{algline:comparemodes} selects eigenvalues that are approximately common to \texttt{modes50} and \texttt{modes80}. As described in the Section \ref{subsec:printcommands}, this may serve as input for the \texttt{PrintTable} command in line \ref{algline:printtable}. We have chosen to rescale the eigenenergies in lines \ref{algline:printfrequencies} and \ref{algline:printtable} so that the output would be the first 10 perfect squares. 

\begin{figure}[!t]
\caption{\label{fig:quickmodes} Output of Notebooks \ref{alg:basiccommands}. Top: (\texttt{PrintFrequencies}) The first 10 eigenenergies calculated using a basis degree of 50, plotted on the complex plane. Middle: (\texttt{PrintEigenfunctions}) The eigenfunctions of the first 3 eigenenergies, calculated using a basis degree of 50, normalized according to their $L^2$-norm. Bottom: (\texttt{PrintTable}) Rescaled eigenvalues common to basis degrees of 50 and 80. There are 28 eigenenergies that share a minimum of 3 significant digits (not shown). We tabulate only the lowest 10. The spectrum calculated is in excellent agreement with \eqref{eq:ISWenergies}.}
\vspace{1em}
%\begin{indented}
%\item[]
\begin{tabular}{c}
\includegraphics[width=0.4\textwidth]{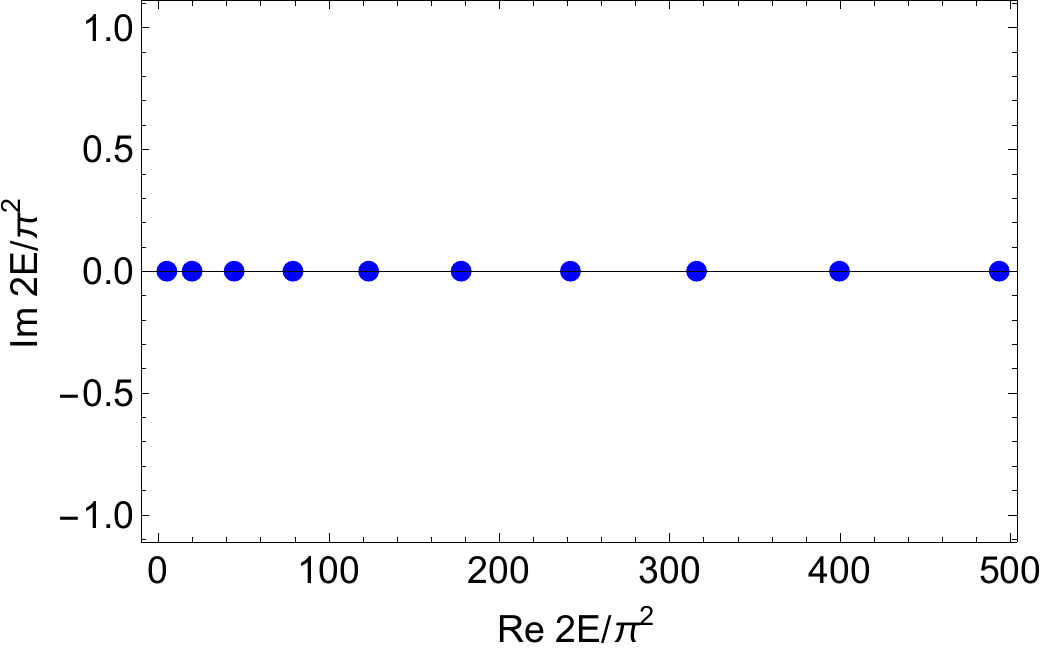} \hspace{28pt} \\
\includegraphics[width=0.4\textwidth]{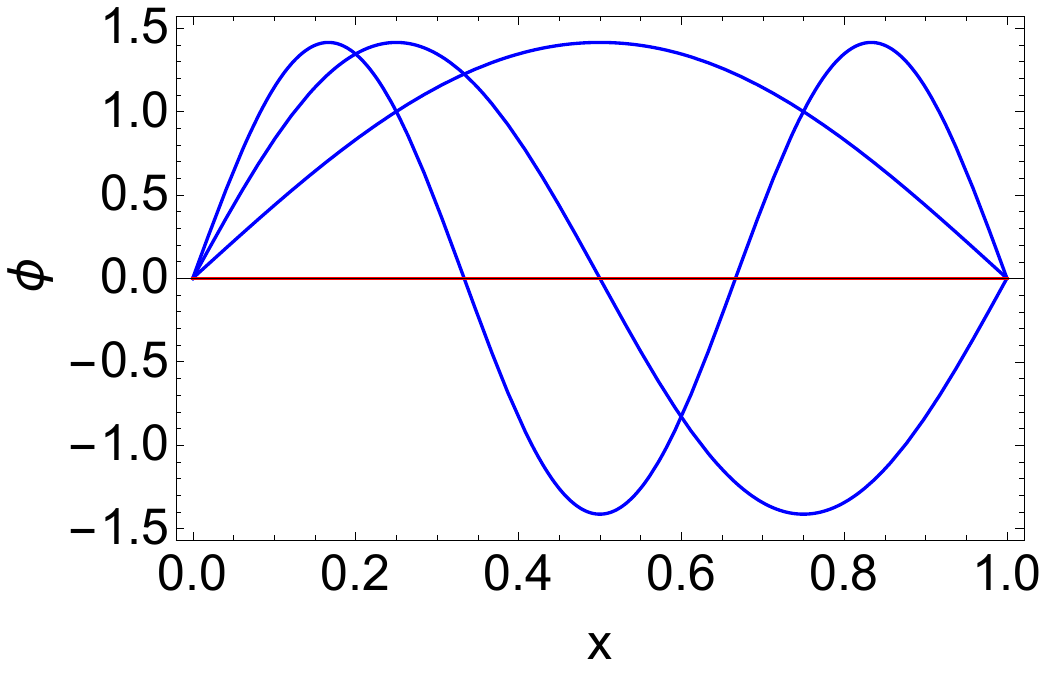} \hspace{20pt} \\
\includegraphics[width=0.3\textwidth]{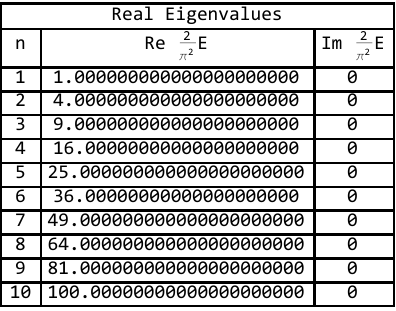}
\end{tabular}
%\end{indented}
\end{figure}

Line \ref{algline:eigenfunctions} plots the eigenfunctions of the inputted spectrum of the lowest three eigenvalues of \texttt{modes50} using a basis degree of $50$. The \texttt{Print} commands found in the last 3 lines output Figure \ref{fig:quickmodes}.

As described in Section~\ref{subsec:printcommands}, the command \texttt{PrintTable} only prints out \textit{significant digits}. As an illustrative example, consider the lowest rescaled eigenenergies. The absolute error for \texttt{modes50} is $3.27\times10^{-22}$ and the absolute error for \texttt{modes80} is $4.97\times10^{-31}$. The \texttt{PrintTable} compares the two eigenvalues and detects a difference of $\sim 10^{-22}$, and prints out the eigenvalue up to the $21^{st}$ decimal place.

\subsubsection{\texttt{SpectralBP}$-$quick commands.}

Three commands can do the calculations in Notebook \ref{alg:basiccommands}. We have omitted the relevant options for boundary conditions and printing for conciseness. Notebook~\ref{alg:quickcommands} outputs the same figures as in Notebook~\ref{alg:basiccommands}.

\begin{algorithm}[H]
\caption{: A shorter implementation of \texttt{SpectralBP} equivalent to Notebook~\ref{alg:basiccommands}}\label{alg:quickcommands}
\begin{algorithmic}[1]
\State TISE =  Equation \eqref{eq:TISE} with potential \eqref{eq:ISWpotential}
\State quickmodes = GetAccurateModes[TISE, 50, 80, ...];
\State PrintAll[TISE,quickmodes,50, ...]
\end{algorithmic}
\end{algorithm}

\subsubsection{A note on machine precision.}
As described in Section \ref{subsec:getcommands}, one may use arbitrary precision numbers by inputting a basis tuple of the form \texttt{\{N,prec\}}. This would calculate eigenvalues using a basis degree of \texttt{N} with \texttt{prec}-precision numbers, as in Notebook \ref{alg:precision}.

\begin{algorithm}[H]
\caption{: An implementation demonstrating the use of arbitrary precision numbers in \texttt{SpectralBP}.}\label{alg:precision}
\begin{algorithmic}[1]
\State TISE =  Equation \eqref{eq:TISE} with potential \eqref{eq:ISWpotential}
\State quickmodes = GetAccurateModes[TISE, \{50,50\}, \{80,80\}, ...];
\State PrintTable[$\dfrac{2}{\pi^2}$quickmodes[[;;,1;;10]], FreqName$\to$`$\dfrac{2}{\pi^2}$E'] \Comment{Figure \ref{fig:InfiniteSquareWellTable2}} \label{algline:printtableprec}
\end{algorithmic}
\end{algorithm}

\begin{figure}[!t]
\begin{center}
\includegraphics[width=0.45\textwidth]{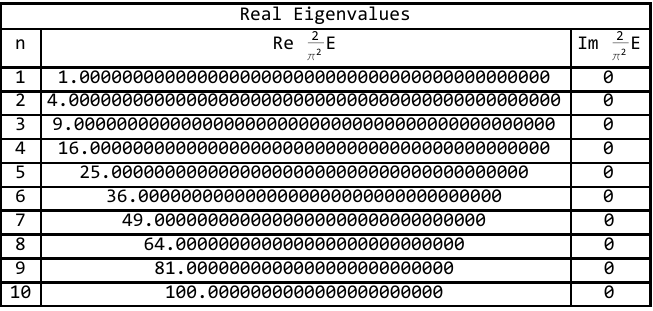}
\caption{Calculated eigenvalues common to basis tuples $\{50,50\}$ and $\{80,80\}$ (described in Section \ref{subsec:getcommands}). There are 28 eigenenergies that share a minimum of 3 significant digits (not shown)$-$similar to Figure \ref{fig:quickmodes}$-$while the number of significant digits for the lower eigenvalues have increased.}
\label{fig:InfiniteSquareWellTable2}
\end{center}
\end{figure}

The \texttt{PrintTable} command in line \ref{algline:printtableprec} outputs Figure \ref{fig:InfiniteSquareWellTable2}. The number of common modes remain at 28 (not shown), but there are more significant digits for the lowest eigenenergies. 

This is because the error due to floating point arithmetic at machine precision is generally small enough to resolve approximately common eigenenergies between basis degrees. When higher precision numbers are used, this error is pushed down further and may reveal more significant digits. The absolute error from approximating the solution space in a finite polynomial basis eventually dominates, and may only be corrected by using higher and higher basis degrees.

Briefly, increasing machine precision increases the significant digits (up to a point) while increasing the Bernstein basis degree used increases the number of converged modes (up to a point).

\subsubsection{Test on non-analytic solutions.}
For completeness, let us explore the case when the exact solution is non-analytic. Suppose we perturb the potential by lifting half of the infinite square well,
\begin{equation}
V(x) = \left\lbrace 
\begin{array}{ll}
0 & 0 \leq x < \dfrac{1}{2} \\
1 & \dfrac{1}{2} \leq x < 1 \\
\infty & \mathrm{otherwise}.
\end{array}
\right.
\label{eq:ISWpertpot}
\end{equation}
The exact solution can be derived by starting with a pair of free particle solutions at $0 \leq x \leq 1/2$ and $1/2 \leq x \leq 1$, then imposing the correct boundary conditions at the walls of the infinite square well and continuity relations $x = 1/2$. One then finds that for the boundary conditions and the continuity relations to be satisfied, the eigenenergies must be solutions to the transcendental equation,
\begin{equation}
\sqrt{2(E-1)} \cot\left( \dfrac{\sqrt{2(E-1)}}{2}\right) + \sqrt{2E} \cot{\dfrac{\sqrt{2E}}{2}} = 0.
\label{eq:ISWtranscendental}
\end{equation}
Exact solutions are non-analytic since they are not twice differentiable at $x = 1/2$.

On the other hand, one may simply swap in the potential \eqref{eq:ISWpertpot} and use a \texttt{GetAccurateModes} to numerically solve for these eigenenergies. We benchmark \texttt{SpectralBP} against the \texttt{Mathematica} in-built function \texttt{NSolve} in Table \ref{table:compare}. \texttt{NSolve} is a zero-finding algorithm, which we use to find solutions to \eqref{eq:ISWtranscendental}. There is great agreement between the two methods. 

The non-analyticity of the solutions has adversely affected how quickly the eigenenergies converge to the correct values, which is expected from a spectral method. \texttt{SpectralBP} was able to find all eigenenergies below 1000. On the other hand, \texttt{NSolve} will not find the eigenenergies indicated by *'s by default. These roots are sensitive since one must start close to the them so that \texttt{NSolve} can find them. The eigenenergies indicated by *'s were found by sampling the range [0,1000] with a resolution of $0.01$.

We note that we have chosen odd basis tuples in the calculation so that the corresponding collocation grids avoids the point $x=1/2$. Choosing even basis tuples degrades the accuracy of odd-numbered eigenenergies, and one would need to reach a basis degree of around 400 to determine the ground state energy accurate to 3 digits. \\

\begin{table}[t]
	\caption{\label{table:compare} Comparison between \texttt{NSolve} (a zero-finding algorithm in \texttt{Mathematica 11.3}) and \texttt{SpectralBP}, for eigenenergies in the range $0 \leq E \leq 1000$. Eigenenergies with *'s were found by \texttt{NSolve} by sampling the range [0,1000] with a resolution of $0.01$. Unmarked eigenenergies can be found by default. Eigenenergies found by \texttt{SpectralBP} used basis tuples of $\{61,61\}$ and $\{101,101\}$ (described in Section \ref{subsec:getcommands}). There is excellent agreement between the eigenenergies found by both methods.} \vspace{1em}
%\begin{indented}
%\item[]
\begin{center}
\begin{tabular}{ccc}
\hline
$n$ & $E\,\,(\texttt{NSolve})$ & $E\,\, (\texttt{SpectralBP})$ \\
\hline
1 & 5.422146460 & 5.4221 \\
2 & 20.24869744 & 20.2487 \\
3 & 44.91181375 & 44.9118 \\
$4^*$ & 79.45920945 & 79.4592 \\
5 & 123.8695486 & 123.8695 \\
$6^*$ & 178.1539346 & 178.1539 \\
7 & 242.3050494 & 242.3050 \\
$8^*$ & 316.3279345 & 316.3279 \\
9 & 400.2188219 & 400.2188 \\
$10^*$ & 493.9806000 & 493.9806 \\
11 & 597.6109616 & 597.6110 \\
$12^*$ & 711.1117807 & 711.1118 \\
13 & 834.4814970 & 834.4815 \\
$14^*$ & 967.7214252 & 967.7214 \\
\hline
\end{tabular}
\end{center}
%\end{indented}
\end{table}

\subsection{Quantum harmonic oscillator}
\label{subsec:QHO}

Consider the harmonic oscillator potential,
\begin{equation}
V(x) = \dfrac{1}{2} x^2.
\label{eq:QHOpotential}
\end{equation}
Its eigenenergies are,
\begin{equation}
E_n = n + \dfrac{1}{2}, \qquad n = 0,1,2,\dots
\label{eq:QHOenergies}
\end{equation}
The domain of the solutions is the entire real line $(-\infty,\infty)$ with boundary conditions
\begin{equation}
\lim_{x \to -\infty} \phi(x) \sim 0, \qquad \lim_{x \to \infty} \phi(x) \sim 0.
\label{eq:HObounds}
\end{equation}

\begin{table}[t]
	\caption{\label{table:HOTISE} Comparison between compactifying using \eqref{eq:HOTISE1change} and \eqref{eq:HOTISE2change}, using Bernstein tuples $\{50,50\}$ and $\{100,100\}$ (described in Section \ref{subsec:getcommands}). For conciseness we indicate eigenergies found using \eqref{eq:HOTISE1change} with a dagger$^\dagger$, and mark in square brackets the additional significant digits calculated using \eqref{eq:HOTISE2change}. Compactifying using \eqref{eq:HOTISE2change} performs better, which finds more eigenvalues with more significant digits.} \vspace{1em}
%\begin{indented}
%\item[]
\begin{center}
\begin{tabular}{cc}
$n$ & $E_n$ \\
\hline \hline
$1^\dagger$ & 0.500000[0000000] \\
$2^\dagger$ & 1.50000[00000000] \\
$3^\dagger$ & 2.500[0000000] \\
4 & 3.500000000 \\
5 & 4.50000000 \\
6 & 5.5000000 \\
7 & 6.500000 \\
8 & 7.50000 \\
9 & 8.50000 \\ 
10 & 9.5000 \\
11 & 10.500 \\
12 & 11.500 \\
13 & 12.50 \\
14 & 13.50 \\
15 & 14.50 \\
\hline \\
\end{tabular}
\end{center}
%\end{indented}
\end{table}

\subsubsection{Compactification and boundary conditions.}

One may swap in the harmonic oscillator potential in the example notebooks we have presented to calculate eigenenergies and eigenfunctions, except one must include an additional step of compactifying the domain. Let us compare the spectrum calculated using two different ways of compactifying the interval $(-\infty,\infty)$. The first,
\begin{equation}
v_1 = \tanh(x), \label{eq:HOTISE1change}
\end{equation}
has a domain of $[-1,1]$. As described in Section \ref{subsec:getcommands}, one may change the default domain of $[0,1]$ to $[-1,1]$ by setting the option value of \texttt{LowerBound} to -1. The second,
\begin{equation}
v_2 = \dfrac{1}{1+ \exp(-x)} \label{eq:HOTISE2change}
\end{equation}
has a domain of $[0,1]$.

\begin{figure}[t]
\begin{center}
\begin{tabular}{c}
\hspace{10pt} \includegraphics[width=0.40\textwidth]{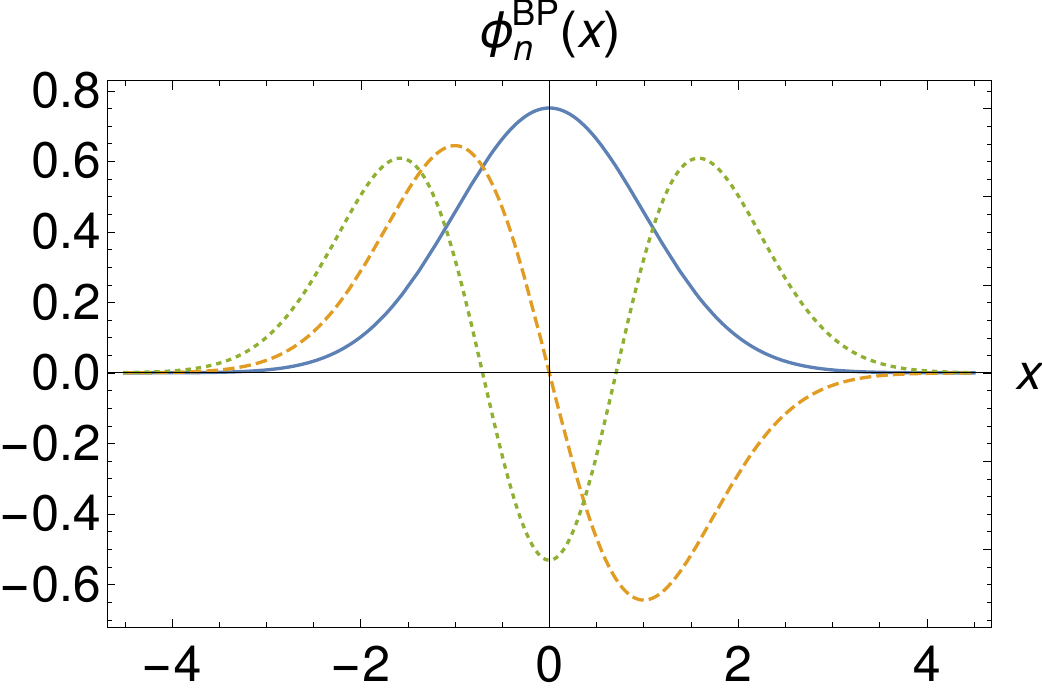} \\
\includegraphics[width=0.40\textwidth]{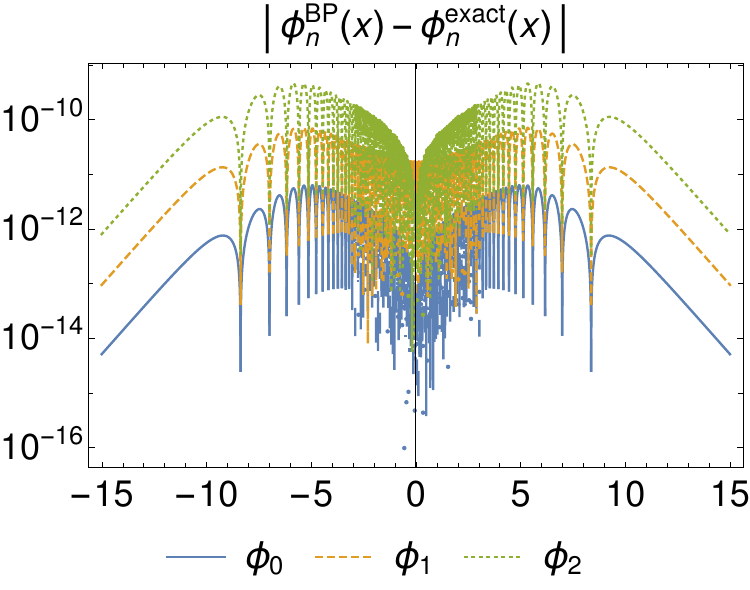} \\
\end{tabular}
\caption{The calculated eigenfunctions $\phi^{\mathrm{BP}}_n(x)$ in the uncompactified coordinate system are plotted above, while the absolute difference between $\phi^{\mathrm{BP}}_n$ and the exact eigenfunctions $\phi^{\mathrm{exact}}_n$ are plotted below. The eigenfunctions were calculated with a basis degree of 100.}
\label{fig:HOTISEFunc}
\end{center}
\end{figure}

Some comments are in order. First, note that the exact solution in both compactified coordinates is flat at both boundaries. All derivatives vanish at either boundary. However, it is sufficient to specify at least 
\begin{equation}
\lim_{v_k \to a_k} \psi(v_k) \sim (v_k-a_k), \quad \lim_{v_k \to b_k} \psi(v_k) \sim (b_k-v_k),
\end{equation}
where $a_k,b_k$ are the corresponding boundary locations for $k=1,2$. Second, note that the potential is singular at the boundaries in both compactified coordinates, with
\begin{equation}
V(v_1) = \dfrac{1}{2} \left(\mathrm{tanh}^{-1}(v_1)\right)^2, \;\; V(v_2) = \dfrac{1}{2} \left( \ln \left(\dfrac{v_2}{1 - v_2}\right) \right)^2.
\end{equation}
A consequence of using the collocation grid we proposed in Section \ref{subsect:LEP} is that we have avoided evaluating at these singularities by expanding the Bernstein basis order and choosing collocation points in the interior of the relevant domain.

Finally, we observe a dependence on the rate of convergence of the method with respect to different coordinate transformations, as can be seen in Table~\ref{table:HOTISE}. We may attribute this discrepancy on how features such as maxima and nodes of higher energy eigenfunctions are distributed on the compactified coordinates in relation to how the collocation points are distributed on the compactified coordinates.

Consider the distance of the right-most maxima or node relative to the upper bound of a high energy eigenfunction for either transformation,
\begin{equation}
    \lim_{x \to \infty} 1 - v_1(x) \sim 2 \exp(-2x), \qquad \lim_{x \to \infty} 1 - v_2(x) \sim \exp(-x).
\end{equation}
That is, in proportion to the length of the interval, these features are closer to the upper bound of the interval with \eqref{eq:HOTISE1change} than in \eqref{eq:HOTISE2change},
\begin{eqnarray}
    \eqref{eq:HOTISE1change} \to \exp(-2x), \qquad \eqref{eq:HOTISE2change} \to \exp(-x)
\end{eqnarray}
The same is true for the lower bound.

Thus, a collocation grid defined on $v_1$ is unable to resolve higher energy eigenfunctions compared to $v_2$ since the collocation points are less densely located on where the maxima and nodes are expected to appear - ie., closer to the edge in proportion to the length of the interval for \eqref{eq:HOTISE1change} than in \eqref{eq:HOTISE2change}

We note that a transformation such as
\begin{equation}
    v_3 = \dfrac{1}{1 + \exp(-x/2)}
\end{equation}
`spreads' these features further away from the upper bound and lower bound. An identical calculation on $v_3$ yields more accurate eigenenergies than on $v_2$ as well as finding more eigenenergies (upto 26).

\subsubsection{Eigenfunctions$-$normalization and manipulation.}

Consider the eigenfunctions calculated from \eqref{eq:HOTISE1change} and \eqref{eq:HOTISE2change}. To properly normalize the eigenfunctions in the original coordinates $x$, one must introduce a weight function underneath the integral of their $L^2$-norms in the compactified coordinates, respectively of the form
\begin{eqnarray}
w(v_1) &=& (v_1 + 1)^{-1} (1 - v_1)^{-1} \\
w(v_2) &=& v_2^{-1} (1 - v_2)^{-1}
\end{eqnarray}
As described in Section \ref{subsec:getcommands}, the option value for \texttt{Normalization} should be \texttt{\{"L2Norm",\{1,-1,-1\}\}} for both $v_1$ and $v_2$.

The eigenfunctions of the three lowest eigenenergies in Table \ref{table:HOTISE} may be calculated using the \texttt{GetEigenfunctions} command. The output is a Bernstein polynomial in the compactified variable $v_2$, which may reverted to the original uncompactified coordinates by a change of variables. The eigenfunctions in $x$ are plotted in Figure \ref{fig:HOTISEFunc} together with their absolute error compared with the exact eigenfunctions. The absolute error is bounded from above, with a maximum difference between $10^{-9}-10^{-11}$.

\subsection{Anharmonic potentials}
\label{subsec:AHO}

\begin{table*}
	\caption{\label{table:APTISE} Spectra for anharmonic potentials found in \eqref{eq:anharmonic1} and \eqref{eq:anharmonic2}, with $\lambda = 1/7$ and $\beta = 40/49$, calculated using basis tuples $\{250,250\}$ and $\{300,300\}$ (described in Section \ref{subsec:getcommands}). Only common eigenvalues with at least 5 significant digits were kept. For \eqref{eq:anharmonic2}, there are 79 such eigenvalues. We have chosen to show only the lowest 10 eigenvalues up to 40 digits, rounded up.}\vspace{1em}
%\begin{indented}
%\item[]
\center
\begin{tabular}{ccc}
$n$ & $E_n, V(x) = \dfrac{1}{4}x^2 + \dfrac{i}{7}x^3$ & $E_n, V(x) = x^2 + \dfrac{40}{49} x^4$\vspace{3pt} \\ 
\hline \hline
1 & 0.6127381063889841 & 1.342244421251821063337113841770966554914 \\
2 & 2.04730063616096 & 4.452375736716380532505970385912143312626 \\
3 & 3.6798624029746 & 8.244544675014299218649219540133247124221 \\
4 & 5.439569424420 & 12.49407778263995078092853450174005121828 \\
5 & 7.2967453569 & 17.11263824817696165379262553962839173473 \\
6 & 9.23400490 & 22.04540267622473136055899649692357072940 \\
7 & 11.2397435 & 27.25459145550393471355991795806437315617 \\
8 & 13.305592 & 32.71221322542317264941304638323745171222 \\
9 & 15.42519 & 38.39651749713872030763192575022745155447 \\
10 & 17.5935 & 44.2900140333829641035044762689148342848 \\
\hline
\end{tabular}
%\end{indented}
\end{table*}

We now benchmark \texttt{SpectralBP} against other numerical methods, here in the context of anharmonic potentials. We perform test calculations also done in \cite{Bender2001a} and \cite{Ezawa2014}, in which the time-independent Schr\"odinger equation has been rescaled such that,
\begin{equation}
\phi''(x) + (E - V(x)) \phi(x) = 0,
\end{equation}
and the anharmonic potentials,
\begin{eqnarray}
V(x) &=& \dfrac{1}{4} x^2 + i \lambda x^3, \label{eq:anharmonic1}\\
V(x) &=& x^2 + \beta x^4 \label{eq:anharmonic2}
\end{eqnarray}
were considered. In the papers cited, Pad\'e approximation and Milne's method \cite{Milne1930} were used to calculate the ground state energies.

The potential \eqref{eq:anharmonic1} is interesting. Although the corresponding Hamiltonian,
\begin{equation}
H = p^2 + \dfrac{1}{4} x^2 + i \lambda x^3,
\end{equation}
isn't hermitian, its eigenenergies remain real and positive. This is because of its underlying $\mathcal{PT}$ symmetry \cite{Bender1998}, in which combining parity, $\mathcal{P}: p \to -p$ and $x \to -x$, and time reversal, $\mathcal{T}: p \to -p,$ $x \to x,$ and $i \to -i$, transformations leaves $H$ invariant.

For both potentials, we compactify our domain via the transformation in \eqref{eq:HOTISE2change}. To recreate Table II of \cite{Bender2001a}, we set $\lambda = 1/7$ and $\beta = 40/49$ and use basis tuples $\{250,250\}$ and $\{300,300\}$ (described in \ref{subsec:getcommands}). The spectra of both potentials are found in Table \ref{table:APTISE}. For a more direct comparison to Table II of \cite{Bender2001a}, we use Equations (8) and (9) of \cite{Bender2001a} to calculate $P(\lambda^2)$ and $P(\beta)$ for the ground state energy. Comparing the two values coming from both basis tuples for significant digits, and we arrive at the expressions
\begin{eqnarray}
P(\lambda^2) &=& 5.524167213060[22] \nonumber \\
P(\beta) &=& 0.41924941603348[0802587964456...]. \nonumber
\end{eqnarray}
where the last expression goes on for 21 more digits. These values are in excellent agreement with the values calculated in \cite{Bender2001a}. The digits enclosed in square brackets are additional significant digits calculated by \texttt{SpectralBP}.

The anharmonic potential \eqref{eq:anharmonic2} was used in \cite{Ezawa2014}, but for different values of $\beta$. We calculated spectra using basis tuples of \{150,150\} and \{200,200\}, keeping only eigenvalues with at least 5 significant digits. In Table \ref{table:APTISE3}, we show only the ground state energies for a direct comparison of Table II and Table IV of \cite{Ezawa2014}.

The results are in great agreement with the ``Exact'' values calculated in \cite{Ezawa2014}, which were calculated using Milne's method \cite{Milne1930}. At the digits where they differ, which we have indicated in square brackets, the difference is within the error bars in both tables. The calculations took an average of 68 seconds each, running in a single 2.50 GHz Intel i5 Core with 8.00GB RAM. 

With modest resources, we are able to calculate the ground state energies to high precision. This is simultaneous with an abundance of excited state energies; the calculation at $\beta = 1/10$ yielded 47 eigenenergies with at least 5 significant digits, while the calculation at $\beta = 100$ yielded 69 eigenenergies with at least 5 significant digits.

\section{Applications in Quasinormal Modes}
\label{sec:QNM}

In general relativity, spacetime itself is treated as a dynamical entity, interacting with the matter that is placed within it. Thus, black holes found in nature are always interacting with complex distributions of matter and fields around them. In active galactic nuclei, accretion disks transport matter inward and transport angular momentum outward, heating the accretion disk into a hot plasma and immersing the black hole in a complex gravitational and electromagnetic system. Even in the absence of matter and fields, the black hole interacts with the vacuum around it, slowly evaporating due to Hawking radiation.

The standard treatment is to decompose the spacetime as in
\begin{equation}
g_{\mu \nu} = g^{0}_{\mu \nu} + \delta g_{\mu \nu},
\label{eq:metricpert}
\end{equation}
where the metric $g^{0}_{\mu \nu}$ is that of an unperturbed black hole, such as the Schwarzschild or Kerr solution. In the linear approximation $\delta g_{\mu \nu} \ll g^{0}_{\mu \nu}$ (so called because the perturbing metric $\delta g_{\mu \nu}$ does not back react with the background metric), these small perturbations generically take the form of damped oscillations known as \textit{quasinormal modes}. When $g^{0}_{\mu \nu}$ is spherically-symmetric, the equations for $\delta g_{\mu \nu}$ reduce to one-dimensional wave equations in certain potentials. These are the famous Regge-Wheeler and Zerilli equations for odd- and even-parity perturbations, respectively.

Quasinormal modes arise as the characteristic ringing of spacetime as it is perturbed by some external field. For a given external field, these oscillations are independent of the initial excitation, their vibrations and damping specified solely by the mass, spin and charge of the black hole. As such, quasinormal modes are used as probes for the validity of general relativity in the strong gravity regime.

From a more theoretical perspective, quasinormal modes provides a test for the linear stability of more exotic spacetimes (such as black branes, black rings, black string): when all quasinormal modes are damped ($\texttt{Im}(\omega) \leq 0$), the spacetime is linearly stable. In the context of AdS/CFT duality, the onset of instability of the AdS spacetime corresponds to a thermodynamic phase transition in CFT.

Review articles on quasinormal modes in an astrophysical setting - black holes, stars, and other such compact objects - we cite \cite{Kokkotas1999} and \cite{Nollert1999}. An emphasis on higher dimensional black holes and their connection to strongly coupled quantum fields is in \cite{Berti2009}, while \cite{Konoplya2011} emphasizes on the various numerical and analytical techniques that have been developed to calculate quasinormal modes. The papers \cite{Grandclement2009,Dias2016} focus on the application of spectral and pseudospectral methods in gravity, of which \texttt{SpectralBP} is an example of.

\begin{table*}
	\caption{\label{table:APTISE3} Ground state energies calculated using the anharmonic potential \eqref{eq:anharmonic2} for different values of $\beta$, using basis tuples $\{150,150\}$ and $\{200,200\}$ (described in Section \ref{subsec:getcommands}). For conciseness, we have enclosed in square brackets additional significant digits calculated by \texttt{SpectralBP} compared to an application of Milne's method in \cite{Ezawa2014}.}\vspace{1em}
%\begin{indented}
%\item[]
\center
\begin{tabular}{cc}
$\beta$ & $E_1$ \\
\hline \hline
0.1 & 1.065285509543717688857091628[8] \\
0.2 & 1.118292654367039153430813153[84] \\
1.0 & 1.392351641530291855657507876[60993418] \\
10 & 2.449174072118386918268793906[187730426220277999] \\
100 & 4.999417545137587829294632037[34965271862550738578] \\
\hline
\end{tabular}
%\end{indented}
\end{table*}

\subsection{Regge-Wheeler equation}
\label{subsec:analytics}
In Section \ref{sec:SBPintro}, we described a general work flow starting from an ODE eigenvalue problem. In this subsection we go through the first 3 steps of this work flow, starting from a standard ODE eigenvalue problem for quasinormal modes. We focus on the Regge-Wheeler equation as an illustrative example; a treatment of the Zerilli equation would proceed in a similar manner. The Regge-Wheeler equation describes axial or odd-parity perturbations of the Schwarzschild metric of mass $M$ linearly coupled to a perturbing field of spin $s$ and angular momentum $l$,
\begin{eqnarray}
\partial_t^2 \Phi(t,r_*) + \left( - \partial_{r_*}^2 + V(r_*^2) \right) \Phi(t,r_*) = 0,\\
V(r_*) = \left(1 - \dfrac{1}{r}\right) \left( \dfrac{l(l+1)}{r^2} + \dfrac{1-s^2}{r^3}\right),
\label{eq:RWEwaveequation}
\end{eqnarray}
where $r_* = r + r_s \ln(r/r_s - 1)$. We are interested in solutions of the form $\Phi(t,r_*) = R(r) \exp(-i \epsilon t)$. This then turns (\ref{eq:RWEwaveequation}) into the ODE eigenvalue problem of the form,
\begin{multline}
\dfrac{\epsilon^2 r^4 - l(l+1)r^2 +(l(l+1) + s^2 - 1)r + 1 - s^2}{r^2 (r-1)^2} R(r) \\ \hspace{6em} + \dfrac{1}{r(r-1)} \deri{R}{r} + \deri{^2 R}{r^2}  = 0,
\label{eq:RWE}
\end{multline}
%\begin{equation}
%\begin{array}{l}
%\deri{^2 R}{r^2} + \dfrac{1}{r(r-1)} \deri{R}{r} \\ \hspace{2em}+ \dfrac{\epsilon^2 r^4 - l(l+1)r^2 +(l(l+1) + s^2 - 1)r + 1 - s^2}{r^2 (r-1)^2} R(r) = 0,
%\end{array}
%\label{eq:RWE}
%\end{equation}
with $\epsilon = 2M \omega$. The domain of the solutions relevant to us is non-compact, stretching from the black hole horizon at $r = 1$ to spatial infinity at $r = \infty$. Note also that the solutions are non-analytic. The coordinate singularity at $r=0$ and the black hole horizon at $r=1$ are both regular singular points of the ODE, while spatial infinity $r=\infty$ is an irregular singular point of the ODE.

We may peel away the non-analytic parts by rescaling out the asymptotic behaviour of $R(r)$ at the black hole horizon and at spatial infinity. The asymptotic behaviour of $R(r)$ at $r = \infty$ can be easily determined to be
\begin{equation}
R^{\mathrm{out}}(r) \sim r^{ i \omega} \exp ( i \omega r) \quad R^{\mathrm{in}}(r) \sim r^{- i \omega} \exp (- i \omega r),
\end{equation}
where we have indicated in superscript which solution is outgoing or ingoing at spatial infinity when the time dependence is restored.

Since the singularity at $r=1$ is regular, we may write an indicial equation $f(x)=0$ at $r = 1$. This can be shown to be simply
\begin{equation}
x^2 + \omega^2 = 0
\label{eq:indicialequation}
\end{equation}
which defines two solutions around $r=1$,
\begin{equation}
R_{\mathrm{in}}(r) \sim (r-1)^{-i \omega} \qquad R_{\mathrm{out}}(r) \sim (r-1)^{i \omega},
\end{equation}
where we have indicated in subscript which solution is outgoing or ingoing at the black hole horizon when the time dependence is restored. 

We expect a perturbation to come from a finite location outside the black hole. As this perturbation propagates, we expect it to either fall into the black hole or out into spatial infinity. This defines the behaviour of the causal solution, and corresponds to the quasinormal mode boundary conditions
\begin{equation}
\lim_{r \to 1} R(r) \sim R_{\mathrm{in}}(r), \qquad \lim_{r \to \infty} R(r) \sim R^{\mathrm{out}}(r).
\end{equation}
An acausal solution would contain parts that are either propagating out of the black hole, or propagating in from spatial infinity. We rescale out the non-analytic parts of the desired solution,
\begin{equation}
R(r) = r^{2 i \omega} (r-1)^{- i \omega} \exp(i \omega r) \phi(r),
\label{eq:RWasymptotics}
\end{equation}
leaving us with a differential equation in $\phi(r)$. We note that the additional factor of $r^{i \omega}$ is there to cancel out the asymptotic behavior of $(r - 1)^{-i \omega}$ around spatial infinity.

Explicitly, the rescaled solution at the boundaries have the following behaviours:
\begin{eqnarray}
\phi_{\mathrm{in}}(r) \sim 1, &\;\;& \phi_{\mathrm{out}}(r) \sim (r-1)^{2 i \omega}, \label{eq:QNMrescaledBH} \\
\phi^{\mathrm{out}}(r) \sim 1, &\;\;& \phi^{\mathrm{in}}(r) \sim r^{- 2 i \omega} \exp(- 2 i \omega r).
\end{eqnarray}
For generic values of $\omega$, these four solutions have very distinct behaviours. Consider the acausal solutions near their corresponding limits,
\begin{eqnarray}
\lim_{r \to 1} \left\lvert \phi_{\mathrm{out}}(r)\right\rvert &=& \left\lbrace
\begin{array}{ll}
\infty, & \qquad \mathrm{Im}~\omega > 0 \\
0, & \qquad \mathrm{Im}~\omega < 0
\end{array} \right. \\
\lim_{r \to \infty} \left\lvert \phi^{\mathrm{in}}(r)\right\rvert &=& \left\lbrace
\begin{array}{ll}
\infty, & \qquad \mathrm{Im}~\omega > 0 \\
0, & \qquad \mathrm{Im}~\omega < 0.
\end{array} \right.
\end{eqnarray}
When $\mathrm{Im}(\omega) = 0$, both solutions are highly oscillatory. Thus, the boundary conditions,
\begin{equation}
\lim_{r \to 1} \phi(r) \sim 1, \qquad \lim_{r \to \infty} \phi(r) \sim 1
\label{eq:qnmboundary}
\end{equation}
filters out both undesired acausal solutions, since these solutions cannot be approximated in the Bernstein basis of finite degree. Thus, with the boundary conditions in \eqref{eq:qnmboundary}, we may identify our solutions to correspond to quasinormal mode eigenfunctions,
\begin{equation}
\phi(r) = \phi^{\mathrm{out}}_{\mathrm{in}}(r).
\end{equation}

\begin{figure}[t!]
\begin{center}
\includegraphics[width=0.45\textwidth]{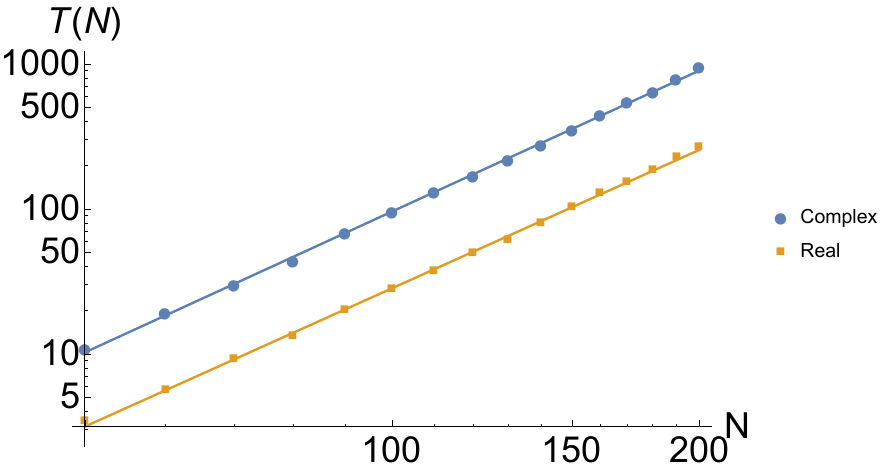}
\caption{Benchmarking for performance using basis tuples $\{N,N\}$. The blue line comes from \eqref{eq:qnmode1}, in which the coefficient functions are complex. The orange line effects the replacement $\omega \to i \lambda$, solving \eqref{eq:qnmode2} in which the coefficient functions are real. Both are power laws of the form $T(N) \sim N^{3.2}$, with the latter performing faster. Calculations were done in a single 2.50 GHz Intel i5 Core with 8.00GB RAM}
\label{fig:qnmbenchmarking}
\end{center}
\end{figure}

Finally, we compactify the region $[1,\infty)$ to $[0,1]$ via the change of variables $r \to 1/u$, leaving us with
\begin{multline}
\left(-l - l^2 + 4 \omega^2 + u (s^2 + (i + 2 \omega)^2)\right) \phi(u) \\
+ (- 2 i \omega + 2 u + u^2 (-3 + 4 i \omega))\phi'(u) \\ - (u-1) u^2 \phi''(u) = 0.
\label{eq:qnmode1}
\end{multline}
%\begin{equation}
%\begin{array}{l}
%\left(-l - l^2 + 4 \omega^2 + u (s^2 + (i + 2 \omega)^2)\right) \phi(u) + \\
%\qquad \qquad (- 2 i \omega + 2 u + u^2 (-3 + 4 i \omega))\phi'(u) - (u-1) u^2 \phi''(u) = 0.
%\end{array}
%\label{eq:qnmode1}
%\end{equation}
This change of variables moves the regular singularity at $r = 0$ to $u = \infty$ and the irregular singularity at $r = \infty$ to $u = 0$.

We may use equation \eqref{eq:qnmode1} as the ODE eigenvalue problem we feed into \texttt{SpectralBP}. However, we may improve our calculations with the transformation $\omega \to i \lambda$, which yields an ODE eigenvalue problem whose coefficient functions are all real,
\begin{multline}
\left(-l - l^2 - 4 \lambda^2 + u (s^2 - (1 + 2 \lambda)^2)\right) \phi(u) \\
+ (2 u - u^2 (3 + 4 \lambda) + 2 \lambda)\phi'(u) \\
- (u-1) u^2 \phi''(u) = 0,
\label{eq:qnmode2}
\end{multline}
%\begin{equation}
%\begin{array}{l}
%\left(-l - l^2 - 4 \lambda^2 + u (s^2 - (1 + 2 \lambda)^2)\right) \phi(u) + \\
%\qquad \qquad (2 u - u^2 (3 + 4 \lambda) + 2 \lambda)\phi'(u) - (u-1) u^2 \phi''(u) = 0,
%\end{array}
%\label{eq:qnmode2}
%\end{equation}
and boundary conditions
\begin{equation}
\lim_{u \to 0} \phi(u) \sim 1, \qquad \lim_{u \to 1} \phi(r) \sim 1
\end{equation}

The spectral matrices constructed from \eqref{eq:qnmode2} are strictly real. This has two consequences. First, the calculation of the spectra is quicker, which is demonstrated in Figure \ref{fig:qnmbenchmarking}. Solving a generalized eigenvalue problem with matrices that are strictly real is computationally cheaper compared when the matrices involved are complex. Second, the calculated eigenvalues come in only two flavours: real eigenvalues, or complex conjugate pairs. Their eigenvectors are similarly real, or come in complex conjugate pairs. 

When we return the imaginary number $i$, the eigenvalues $\omega$ are expected to be strictly imaginary or come in pairs satisfying $\omega = - \omega^*$. In the proceeding subsections, we calculate all eigenvalues and eigenfunctions using \eqref{eq:qnmode2}, and then multiplying the resulting spectra with $i$ to retrieve the spectrum of \eqref{eq:qnmode1}.

\subsection{Scalar perturbations}

\begin{table*}
\begin{minipage}{\textwidth}
\begin{algorithm}[H]
\caption{QNMS - Scalar perturbations}\label{alg:qnmsscalar}
\begin{algorithmic}[1]
\State scalarode =  Equation \eqref{eq:qnmode2} with $s = 0$ and $l = 3$
\State modes50 = GetModes[scalarode, \{50, 50\}] \label{algline:qnmgetmodes1}
\State modes80 = GetModes[scalarode, \{80, 80\}] \label{algline:qnmgetmodes2}
\State modes100 = GetModes[scalarode, \{100, 100\}] \label{algline:qnmgetmodes3}
\State PrintFrequencies[$i\times$modes50] \Comment{output in Figure \ref{fig:ScalarFreq1}} \label{algline:qnmprintfreq}
\State modes5080 = CompareModes[$i\times$modes50, $i\times$modes80] \label{algline:qnmcompare1}
\State PrintTable[modes5080] \Comment{output in Table \ref{table:scalar5080} (a)} \label{algline:qnmprinttable1}
\State modes5080100 = CompareModes[\{$i\times$modes50, $i\times$modes80, $i\times$modes100\}] \label{algline:qnmcompare2}
\State PrintTable[modes5080100[[1;;2]]] \Comment{output in Table \ref{table:scalar5080} (b)} \label{algline:qnmprinttable2}
\State imagmodes = purely imaginary modes of modes5080 \label{algline:imag}
\State testedimagmodes = CompareEigenfunctions[scalarode, $\dfrac{\mathrm{imagmodes}}{i}$, \{50,80\}] \label{algline:qnmcompare3}
\end{algorithmic}
\end{algorithm}
\end{minipage}
\end{table*}

We now calculate the quasinormal modes of a scalar perturbation ($s=0$) for $l=3$. A simple \texttt{Mathematica} implementation is in Notebook \ref{alg:qnmsscalar}.

\begin{figure}[t]
\begin{center}
\includegraphics[width=0.45\textwidth]{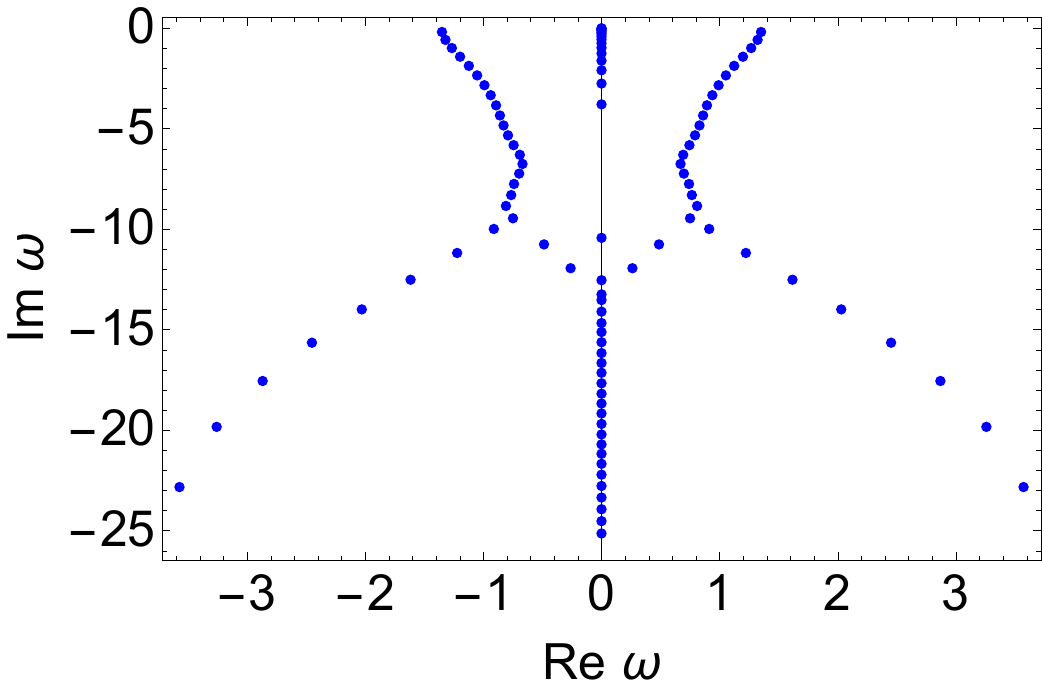}
\caption{Calculated spectrum of a scalar field in a Schwarzschild spacetime for $l=3$ using the basis tuple \{50,50\} (described in Section \ref{subsec:getcommands}), many of which are spurious. There are eigenvalues distributed along the negative-imaginary axis because of the existence of a continuum of eigenvalues that is present there.}
\label{fig:ScalarFreq1}
\end{center}
\end{figure}

The spectrum derived from using a basis tuple of \{50, 50\} (described in Section \ref{subsec:getcommands}) is plotted on the complex plane in Figure \ref{fig:ScalarFreq1}. Since the ODE eigenvalue problem is quadratic in $\omega$, there are 102 eigenvalues as follows from the discussion in Section \ref{subsect:PEP}. 

\subsubsection{Filtering spurious modes.}

\begin{table}[t]
\caption{\label{table:scalar5080} Result of a \texttt{CompareModes} command on 2 and 3 basis tuples (discussed in Section \ref{subsec:getcommands}). (a) The filtered spectrum for the duo basis tuples include purely imaginary modes, which we know to be spurious. These modes may be filtered out using a \texttt{CompareEigenfunctions} command. (b) The filtered spectrum for the trio of basis tuples do not include purely imaginary modes. We have printed here significant digits shared by basis tuples \{80,80\} and \{100,100\}}
\vspace{1em}
%\begin{indented}
%\item[]
\begin{center}
\begin{tabular}{ccc}
\multicolumn{3}{c}{\texttt{Imaginary modes}} \\
\hline \hline
$n$ & Re $\omega_n$  & Im $\omega_n$\\
\hline
1 & $0$ & -18.67 \\
2 & $0$ & -20.70 \\
3 & $0$ & -22.21 \\
\hline
\end{tabular}
\begin{tabular}{ccc}
\multicolumn{3}{c}{\texttt{Complex modes}} \\
\hline \hline
$n$ & Re $\omega_n$  & Im $\omega_n$\\
\hline
1 & $\pm$ 1.35073246507324 & -0.192999255468019 \\
2 & $\pm$ 1.32134299591192 & -0.58456957027682 \\
3 & $\pm$ 1.26725161539 & -0.992016460806 \\
4 & $\pm$ 1.19754651 & -1.42244241 \\
5 & $\pm$ 1.1232546 & -1.877186 \\
6 & $\pm$ 1.05310 & -2.35207 \\
7 & $\pm$ 0.9913 & -2.8408 \\
8 & $\pm$ 0.938 & -3.338 \\
\hline
\end{tabular}
\\ \vspace{4pt} (a) \{50,50\} and \{80,80\} \vspace{8pt} \\
\begin{tabular}{ccc}
\multicolumn{3}{c}{\texttt{Complex modes}} \\
\hline \hline
n & Re $\omega$  & Im $\omega$\\
\hline
1 & $\pm$ 1.35073246507324 & -0.192999255468019 \\
2 & $\pm$ 1.32134299591192 & -0.584569570276824 \\
3 & $\pm$ 1.26725161538865 & -0.992016460806254 \\
4 & $\pm$ 1.1975465055999 & -1.422442414743 \\
5 & $\pm$ 1.1232545798 & -1.8771856473 \\
6 & $\pm$ 1.05309960 & -2.35206873 \\
7 & $\pm$ 0.991268 & -2.840790 \\
8 & $\pm$ 0.93841 & -3.33793 \\
\hline
\end{tabular}
\\ \vspace{4pt} (b)  \{50,50\}, \{80,80\} and \{100,100\}
\end{center}
%\end{indented}
\end{table}

In Section \ref{subsec:comparecommands}, we described two ways to filter out spurious eigenvalues: the \texttt{CompareModes} command and the \texttt{CompareEigenfunctions} command. In Section \ref{sec:TISE}, the \texttt{CompareModes} command on a pair of spectra was sufficient to filter out all the spurious modes. 

In the current case the \texttt{CompareModes} command at line \ref{algline:qnmcompare1} is not sufficient. Its output in Table \ref{table:scalar5080} (a) includes purely imaginary modes, which are well-known not to exist for scalar perturbations given the boundary conditions we have chosen \cite{ferrari1984}.

Recall that equation \eqref{eq:RWE} comes from choosing a stationary ansatz for \eqref{eq:RWEwaveequation}. It has been shown that the retarded Green function of this wave equation possesses a branch cut on the negative-imaginary axis \cite{Leaver1986,Leaver1988}. It is the `shadow' of this continuum of eigenvalues which \texttt{SpectralBP} feels, as can be observed in Figure \ref{fig:ScalarFreq1}.

To filter these modes out, we demonstrate two solutions in the Notebook \ref{alg:qnmsscalar}. These can be found in lines \ref{algline:qnmcompare2} and \ref{algline:qnmcompare3}.

The first method is straightforward: calculate the spectrum of a third basis tuple and select eigenvalues common to all three spectra. We have chosen \{100,100\} as our third basis tuple, and the corresponding output is in Table \ref{table:scalar5080} (b). The purely imaginary modes are successfully filtered out.

The second method would be to compare eigenfunctions between two basis tuples. This is the purpose of the \texttt{CompareEigenfunctions} command, whose output on line \ref{algline:qnmcompare2} is an empty set. This confirms that these modes are indeed spurious; their eigenfunctions are not approximately equal. One is then justified to filter out the purely imaginary modes in Table \ref{table:scalar5080} (a).

The calculation of a third spectrum may be numerically prohibitive, especially when only a small subset of eigenvalues are suspected to be spurious. This consideration would favour one method over the other. In this case testing only the eigenfunctions of the suspected spurious eigenvalues, as filtered in line \ref{algline:imag}, should be favoured over the former method.

This second filter works because the rescaling in equation \eqref{eq:RWasymptotics} keeps other valid solutions of our ODE eigenvalue problem non-analytic. In the case of the branch cut eigenvalues, their corresponding eigenfunctions remains singular at the cosmologcal horizon after rescaling \cite{Casals2013}. Thus, the approximation of these eigenfunctions in a Bernstein basis would fail to converge near the cosmological horizon. This idea is explored further in Section \ref{subsec:rescaling}.

\begin{figure}[b]
\begin{centering}
\begin{tabular}{c}
  \includegraphics[width=0.45\textwidth]{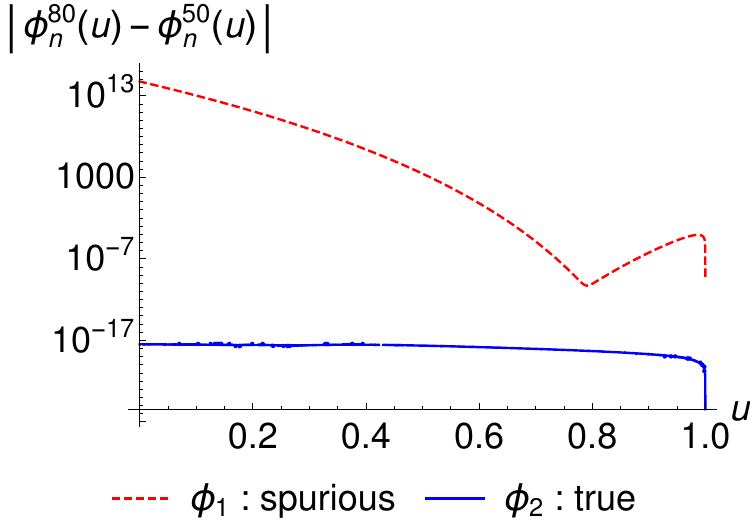} \\
\end{tabular}
\caption{The absolute difference between eigenfunctions of approximately equal eigenvalues using Bernstein basis orders 50 and 80. $\phi_1(u)$ calculates the absolute difference for the eigenvalue $\omega = -18.67 i$, while $\phi_2(u)$ calculates the absolute difference for the eigenvalue $\omega = \pm 1.3507\dots$ $-0.1930\dots i$. The former indicates that the eigenfunctions does not converge to some non-singular function, while the latter indicates convergence.}
\label{fig:QNMerror}
\end{centering}
\end{figure}

This failure to converge is shown explicitly in Figure \ref{fig:QNMerror}, where we compare the eigenfunctions of the spurious eigenvalue $-18.67 i$ and the non-spurious eigenvalue $\pm 1.3507\dots$ $-0.1930\dots i$. 

Using a \texttt{GetEigenfunctions} command, we plotted the absolute difference between the eigenfunctions of approximately common eigenvalues for two spectral basis orders. The maximum error for the spurious eigenvalue is indicative of the presence of a singularity in the eigenfunction,
\begin{eqnarray}
||\phi^{80}_1(u) - \phi^{50}_1(u)||_\infty &\sim& 10^{14}, \nonumber\\
||\phi^{80}_2(u) - \phi^{50}_2(u)||_\infty &\sim& 10^{-17}. \nonumber
\end{eqnarray}

\subsubsection{On the discrete spectrum condition.}
\label{subsec:rescaling}
\begin{figure}[t]
\begin{center}
\includegraphics[width=0.45\textwidth]{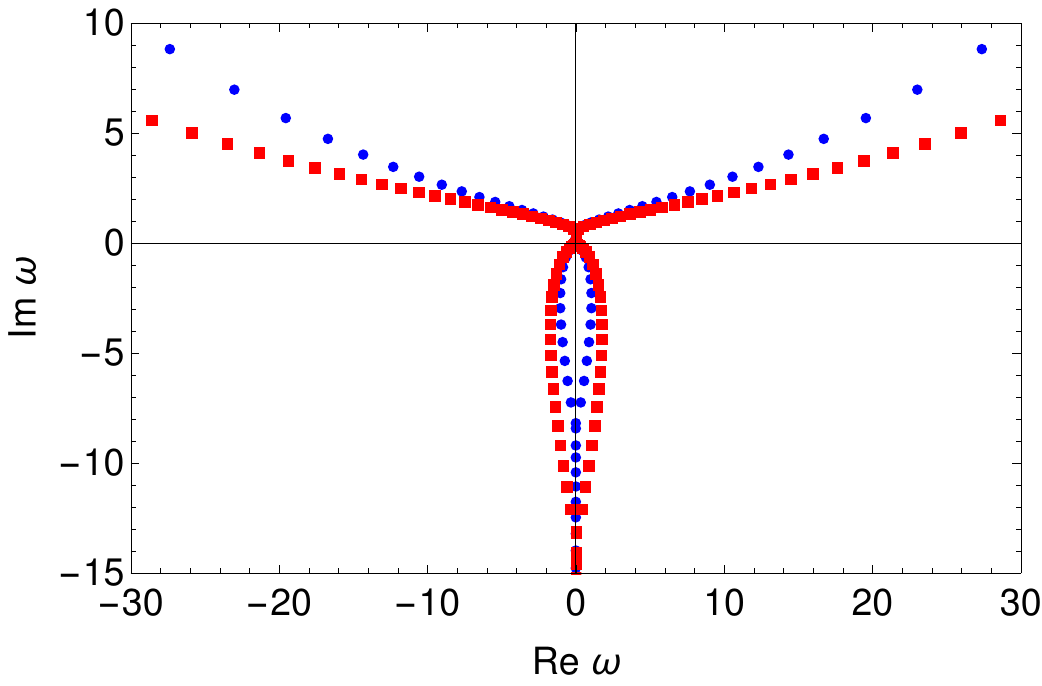}
\caption{Spectrum calculated when $\phi(r)$ is rescaled so that $\lim_{r \to 1} \phi(r) \sim (1-r)$, for basis tuples \{50, 50\} (blue circles) and \{80, 80\} (red squares). The problem has become ill-posed since the rescaling no longer imposes the correct boundary conditions corresponding to a discrete spectrum.}
\label{fig:illposedfreq}
\end{center}
\end{figure}

We echo an idea from \cite{Jansen2017}. One must be careful in rescaling so that boundary conditions are still capable of the undesired solutions. For example, there are instances when peeling off an extra $(r-1)^{-1}$ term so that $\phi(r) \sim (r-1)$ is desirable. This boundary condition would fail to filter out the acausal solution at the black hole horizon, since both the acausal and causal solutions vanish at $r = 1$. The spectral method would then try to solve for solutions of the form,
\begin{equation}
\phi(r) = A \phi^{\mathrm{out}}_{\mathrm{in}}(r) + B \phi^{\mathrm{out}}_{\mathrm{out}} (r),
\end{equation}
which generally is a mixture of causal and acausal parts at the black hole horizon. The ultimate consequence is that the boundary-value problem no longer has a discrete spectrum of eigenvalues. Continuing to calculate the spectrum using \{50, 50\} and \{80, 80\} would result in Figure \ref{fig:illposedfreq}. As expected, \texttt{SpectralBP} is unable to find the desired discrete spectrum.

\section{Algebraically special modes}
\label{sec:ASM}
It is well-known that the standing wave equation for odd- and even-parity gravitational perturbations ($s=2$) has an exact solution at what is called by Chandresekhar as the algebraically special mode. It is a purely imaginary frequency which appears to separate two different branches of the quasinormal mode spectrum: a lower branch that spirals towards the imaginary axis and an upper branch corresponding to an asymptotic high-damping regime.

It is a curious mode, whose frequencies can be shown analytically \cite{Wald1973,Liu1996,Chandrasekhar2006} to be
\begin{equation}
M\omega_l = -i \dfrac{(l-1)l(l+1)(l+2)}{12},
\label{eq:asm}
\end{equation}
and whose corresponding eigenfunctions, with singularities properly scaled out, can be expressed analytically as a truncated polynomial. For example, for $l = 2$,
\begin{eqnarray}
\phi_2(u) = 
1 + \dfrac{115}{7} (u-1) + \dfrac{860}{7} (u-1)^2 + \dfrac{11572}{21} (u-1)^3 + \nonumber \\ 
\dfrac{34486}{21} (u-1)^4 + \dfrac{356662}{105} (u-1)^5 + \dfrac{44372}{9} (u-1)^6 + \nonumber \\
+ \dfrac{44372}{9} (u-1)^7 + \dfrac{77651}{27} (u-1)^8 + \dfrac{11093}{9} (u-1)^9 \qquad
\label{eq:ASM2exact}
\end{eqnarray}
Various numerical investigations \cite{Andersson1994,Leaver2006} are hard-pressed to converge towards this exact result. It has been argued \cite{Liu1996} that the discrepancy can be traced to two explanations: (1) the algebraically special mode is sensitive to the exact form of the gravitational potential (affecting WKB and P\"{o}schl-Teller potential fitting) and (2) the sensitivity of a method to a properly defined mode number (affecting the continued fraction methods by Leaver).

In fact, numerical methods that are able to find eigenvalues on the complex plane do not generally work when those eigenvalues are located exactly on the imaginary axis. For example, the continued fraction method is not convergent for modes on the imaginary axis \cite{Leaver2006,Cook2014,Cook2016}. This disputes previous analytic and numerical results concerning Kerr QNMs on the negative-imaginary axis. One can, however, deduce the existence of these modes by finding `mode sequences' that arbitrarily get close to the negative-imaginary axis, including the special algebraic mode \cite{Berti2003,Cook2014}. How these modes move around the negative-imaginary axis is not accessible to Leaver's method.

With respect to this, spectral methods enjoy a significant advantage over Leaver's method: an algorithm such as \texttt{SpectralBP} is capable of finding eigenvalues on the imaginary axis. Unlike Leaver's method, which is based on a local power series expansion at one of the horizons, spectral methods find solutions globally. This has been reported before in \cite{Jansen2017}, where the spectral algorithm \texttt{QNMspectral} finds a novel infinite set of purely imaginary modes for massless scalar perturbations in a Schwarzschild-de Sitter background. Because the spectral method is able to find these overdamped modes, one is able to observe complex bifurcation events in which quasinormal modes sink into, move along and emerge out of the negative imaginary axis where two QNMs collide. We have also used \texttt{SpectralBP} to uncover an interesting scenario that occurs in a Schwarzschild AdS background \cite{fortuna2022}.

\begin{table}[t]
	\caption{\label{table:qnmgrav1} Gravitational perturbations with $l=2$ and $l=3$, calculated using basis tuples \{350,350\} and \{400,400\}. The special algebraic modes have 295 and 227 significant digits respectively. In units where the horizon is at $r=1$, we have $M=1/2$, so that $\omega_2 = -4i$ and $\omega_3 = -20i$ according to (\ref{eq:asm}). Our numerical results show agreement up to 295 and 227 significant digits, respectively.} \vspace{1em}
%\begin{indented}
%\item[]
\begin{center}
\begin{tabular}{ccc}
\multicolumn{3}{c}{Schwarzschild: $s = 2, l = 2$} \\ [0.5ex] 
\hline \hline
\multicolumn{3}{c}{\texttt{Damped Modes}} \\
\hline 
$n$ & Re $\omega_n$ & Im $\omega_n$ \\
\hline 
0 & $\pm$ 0.747343368836084 & -0.177924631377871 \\
1 & $\pm$ 0.693421993758327 & -0.547829750582470 \\
2 & $\pm$ 0.602106909224733 & -0.956553966446144 \\
3 & $\pm$ 0.503009924371181 & -1.41029640486699 \\
4 & $\pm$ 0.415029159626 & -1.8936897817327 \\
5 & $\pm$ 0.33859881 & -2.39121611 \\
6 & $\pm$ 0.2665046 & -2.895821 \\
7 & $\pm$ 0.1856 & -3.4077 \\
8 & $\pm$ 0.1268 & -4.606 \\
9 & $\pm$ 0.174 & -6.64 \\
\hline \hline
\multicolumn{3}{c}{\texttt{Algebraically Special Mode}} \\
\hline
$n$ & Re $\omega_n$ & Im $\omega_n$ \\
\hline
0 & 0 & -4.\`{}295\\
\hline
\end{tabular}
\\ \vspace{8pt}
\begin{tabular}{ccc}
\multicolumn{3}{c}{Schwarzschild: $s = 2, l = 3$} \\ [0.5ex] 
\hline \hline
\multicolumn{3}{c}{\texttt{Damped Modes}} \\
\hline 
$n$ & Re $\omega_n$ & Im $\omega_n$ \\
\hline
0 & $\pm$ 1.19888657687498 & -0.185406095889895 \\
1 & $\pm$ 1.16528760606660 & -0.562596226870088 \\
2 & $\pm$ 1.10336980155690 & -0.958185501933924 \\
3 & $\pm$ 1.02392382211667 & -1.38067419193848 \\
4 & $\pm$ 0.94034801163031 & -1.83129878501019 \\
5 & $\pm$ 0.86277295728431 & -2.30430272428181 \\
6 & $\pm$ 0.79531904835151 & -2.79182448544518 \\
7 & $\pm$ 0.73798455177946 & -3.28768905671353 \\
8 & $\pm$ 0.689236637190 & -3.78806560839 \\
9 & $\pm$ 0.6473662632 & -4.2907978995 \\
\hline \hline
\multicolumn{3}{c}{\texttt{Algebraically Special Mode}} \\
\hline
$n$ & Re $\omega_n$ & Im $\omega_n$ \\
\hline
0 & 0 & -20.\`{}227 \\
\hline
\end{tabular}
\end{center}
%\end{indented}
\end{table}

\begin{table}[t]
	\caption{\label{table:qnmgrav2} Gravitational perturbations with $l=4$ and $l=5$, calculated using basis tuples \{350,350\} and \{400,400\}. The special algebraic modes have 137 and 115 significant digits respectively. In units where the horizon is at $r=1$, we have $M=1/2$, so that $\omega_4 = -60i$ and $\omega_5 = -140i$ according to (\ref{eq:asm}). Our numerical results show agreement up to 137 and 115 significant digits, respectively.} \vspace{1em}
%\begin{indented}
%\item[]
\begin{center}
\begin{tabular}{ccc}
\multicolumn{3}{c}{Schwarzschild: $s = 2, l = 4$} \\ [0.5ex] 
\hline \hline
\multicolumn{3}{c}{\texttt{Damped Modes}} \\
\hline 
$n$ & Re $\omega_n$ & Im $\omega_n$ \\
\hline 
0 & $\pm$ 1.61835675506448 & -0.188327921977846 \\
1 & $\pm$ 1.59326306406901 & -0.568668698809681 \\
2 & $\pm$ 1.54541906521342 & -0.959816350242326 \\
3 & $\pm$ 1.47967346001108 & -1.36784863803576 \\
4 & $\pm$ 1.40303101850333 & -1.79647794351833 \\
5 & $\pm$ 1.32314499871400 & -2.24595350702581 \\
6 & $\pm$ 1.24621774933184 & -2.71337253668641 \\
7 & $\pm$ 1.17581765005953 & -3.19434136122692 \\
8 & $\pm$ 1.11314953294602 & -3.68463526728615 \\
9 & $\pm$ 1.05799479590004 & -4.18098245812595 \\
\hline \hline
\multicolumn{3}{c}{\texttt{Algebraically Special Mode}} \\
\hline
$n$ & Re $\omega_n$ & Im $\omega_n$ \\
\hline
0 & 0 & -60.\`{}137 \\
\hline
\end{tabular}
\\ \vspace{8pt}
\begin{tabular}{ccc}
\multicolumn{3}{c}{Schwarzschild: $s = 2, l = 5$} \\ [0.5ex] 
\hline \hline
\multicolumn{3}{c}{\texttt{Damped Modes}} \\
\hline 
$n$ & Re $\omega_n$ & Im $\omega_n$ \\
\hline
0 & $\pm$ 2.02459062427070 & -0.189741032163219 \\
1 & $\pm$ 2.00444205578112 & -0.571634763544526 \\
2 & $\pm$ 1.96539152161688 & -0.960656912028150 \\
3 & $\pm$ 1.91000801223541 & -1.36111381729921 \\
4 & $\pm$ 1.84216368773741 & -1.77639518477683 \\
5 & $\pm$ 1.76667152139505 & -2.20836492793496 \\
6 & $\pm$ 1.68849633364143 & -2.65699396530772 \\
7 & $\pm$ 1.61183056559873 & -3.12056171442603 \\
8 & $\pm$ 1.53951216802968 & -3.59636155142390 \\
9 & $\pm$ 1.47299341464745 & -4.08149414445982 \\
\hline \hline
\multicolumn{3}{c}{\texttt{Algebraically Special Mode}} \\
\hline
$n$ & Re $\omega_n$ & Im $\omega_n$ \\
\hline
0 & 0 & -140.\`{}115 \\
\hline
\end{tabular}
\end{center}
%\end{indented}
\end{table}

\subsection{Algebraically special eigenvalues}

We now solve \eqref{eq:qnmode2} for $s = 2$ and for $l = 2,3,4,5$, and reverse the transformation $\omega \to i \lambda$ to retrieve the eigenvalues of \eqref{eq:qnmode1}. We have used basis tuples of \{350,350\} and \{400,400\} (described is Section \ref{subsec:getcommands}) for all calculations, and we have filtered out spurious eigenvalues on the negative-imaginary axis using \texttt{CompareEigenfunctions}. The resulting spectra can be seen in Table \ref{table:qnmgrav1} and Table \ref{table:qnmgrav2}. We show only the 10 lowest damping eigenvalues, using \texttt{Mathematica}'s notation for significant digits for the purely imaginary eigenvalues.

The coincidence of the calculated numerically purely imaginary mode $\omega_l'$ with the algebraically special mode $\omega_l$ is very strong. The coincidence when calculating $\omega_2' \approx \omega_2 = - 4 i$ is within 295 significant digits, $\omega_3' \approx \omega_3 = -20 i$ to within 227 significant digits, $\omega_4' \approx \omega_4 = - 60i$ to within 137 significant digits and $\omega_5' \approx \omega_5 = - 140i$ to within 115 significant digits. This is expected, since we are using a polynomial basis to numerically find a solution whose exact form is a truncated polynomial.

As a additional check, we have verified that the eigenfunction solved by \texttt{SpectralBP} using a basis tuple of \{400,400\} and $l = 2$ is found consistent with \eqref{eq:ASM2exact} to within and error of $10^{-250}$. The eigenfunctions for $l = 3,4$ are also truncated polynomials, of expected degrees 41 and 121 respectively. One might need the use of higher precision numbers to confirm that the degree of the $l=5$ eigenfunction is of degree 281.

As we have described in Section \ref{subsec:analytics}, the eigenvalues of \eqref{eq:qnmode2} are either purely real or come in complex conjugate pairs. As a consequence of this, when we transform back to $\omega$ from $\lambda$ the calculated purely imaginary eigenvalues have exactly no real part. This avoids a criticism on numerical calculations which finds a single mode near the ASM with a finite real part whose symmetric pair $\omega = - \omega^*$ is unexpectedly not found.

The main lesson here is that \texttt{SpectralBP} manages exceptionally well to find eigenvalues on the negative-imaginary axis while filtering out spurious overdamped modes, as would other spectral or pseudospectral methods. This is in contrast with continued fraction methods, which cannot converge when the real part of the eigenvalue vanishes.

As a final note, and to illustrate the resources required to calculate one of the tables in this section, a single spectrum calculation for a basis tuple of $\{400,400\}$ takes around 1 hour each, running in a single 2.50 GHz Intel i5 Core with 8.00GB RAM.

\subsection{Boundary behavior of the eigenfunctions}

For completeness, we give warning when labelling solutions found by spectral methods as bonafide quasinormal modes whenever the eigenvalues calculated imply that the indicial equation \eqref{eq:indicialequation} at one or more of the singularities are non-generic. This may affect whether or not the solution found satisfies the quasinormal mode boundary conditions.

For example, the finiteness of the eigenfunctions of the special algebraic modes at the boundaries can be folded back into \eqref{eq:RWasymptotics}, seemingly then implying that the quasinormal mode boundary conditions are satisfied and that these imaginary frequencies correspond to quasinormal modes.

The indicial equation \eqref{eq:indicialequation} is said to be generic when its two solutions, $\pm i \omega$, do not differ by an integer. This is manifestly true for general complex values of $\omega$. In this case, the power series expansion at $u = 1$ of the rescaled function $\phi(r)$ in \eqref{eq:RWasymptotics} converges, whether dominant or subdominant. At the algebraically special mode, however, the indicial equation is non-generic. From \eqref{eq:asm} and $M = 1/2$, the solution of the indicial equation are both integers,
\begin{equation}
\pm i \omega_l = \dfrac{(l-1)l(l+1)(l+2)}{6}.
\end{equation}
In this case, only one power series expansion of $\phi(r)$ is assured to converge, corresponding to the dominant solution. For the subdominant, say $\tilde{\phi}(r)$, two things may happen. First, the subdominant solution may diverge logarithmically, of the form
\begin{equation}
\tilde{\phi}(u) \sim c_0 \phi(u) \ln (u-1) + a_0 (u-1)^8 + a_1 (u-1)^9 + \dots
\end{equation}
However, a \textit{miraculous} cancellation may occur \cite{MaassenvandenBrink2000}, in which case the logarithmic term vanishes. Thus, both solutions may be expressed as a power series expansion at $r = 1$. It is this latter case that occurs at the algebraically special mode for the Regge-Wheeler equation. This means that the dominant and subdominant solutions, corresponding to ingoing and outgoing modes at the black hole horizon respectively, may be rescaled to have the form,
\begin{equation}
\begin{array}{l}
\phi_{\mathrm{in}}(u) \sim b_0 + b_1 (u-1) + \dots \\
\phi_{\mathrm{out}}(u) \sim a_0 (u-1)^8 + a_1 (u-1)^9 + \dots
\end{array}
\end{equation}

For the specific case of the ASM, the following two statements are then not mutually exclusive: (1) the ASM eigenfunction, properly rescaled, has a regular, well-behaved Frobenius expansion in powers of $(u-1)$ and (2) it is an inextricable mixture of the two linearly independent solutions at the black hole horizon, corresponding to a causal ingoing mode and an acausal outgoing mode. The reconciliation between the analytic and numerical results is thus simple but subtle; \emph{there is no contradiction}. While \texttt{SpectralBP} has indeed found an eigenvalue-eigenfunction pair of the Regge-Wheeler equation, this solution is an inseperable mixture of both ingoing and outgoing solutions at the black hole horizon, and therefore is not a quasinormal mode.

In summary, \texttt{SpectralBP} picks up the special algebraic frequency to an incredible degree of accuracy, but because of the peculiar nature of the special algebraic mode, the corresponding eigenfunction is one that does not satisfy quasinormal mode boundary conditions, as would be expected from \cite{MaassenvandenBrink2000}.

\section{Conclusion}

This work makes a case for the use of Bernstein polynomials as a basis for spectral and pseudospectral methods applied to ordinary differential eigenvalue problems. A prime example of these problems is the calculation of quasinormal modes in black hole spacetimes. The Bernstein polynomials constitute a non-orthogonal spectral basis, which may explain why they are much less utilized compared to Chebyshev or Fourier basis functions. In contrast to its more popular counterparts though, a Bernstein basis allows one to decouple some of the spectral weights relevant to boundary conditions of ordinary differential eigenvalue problems. More specifically, the weights for the first $q$ and last the $r$ basis polynomials for free without recourse to the differential equations. For some applications, this proves to be a significant advantage. 

We developed a user-friendly \texttt{Mathematica} package, \texttt{SpectralBP}, as a general spectral solver for eigenvalue problems. This package fully utilizes the properties of Bernstein polynomials and several other algorithmic enhancements (such as a novel inverse iteration method) that we shall describe in a later paper. As far as we know, \texttt{SpectralBP} is unique among existing spectral codes in its use of a Bernstein basis. We described its key functionalities and showcased several examples for its use. In particular, to serve both as tutorial and benchmarks, we featured applications of \texttt{SpectralBP} to a number of model eigenvalue problems in quantum mechanics. Most importantly, we have also applied \texttt{SpectralBP} to quasinormal mode problems in the Schwarzshild geometry. In all of our example cases, \texttt{SpectralBP} succeeded in providing very accurate results. Remarkably, with only modest resources, we are able calculate the algebraically special modes of Schwarzschild gravitational perturbations. Purely imaginary modes are notoriously difficult to calculate with more conventional numerical methods even when the solution is straightforward to calculate analytically, as in the case for the Schwarzschild ASM. To the best of our knowledge, ours is the most accurate numerical calculation of these algebraically special modes in the extant literature, agreeing with the analytical prediction to a staggering (294!) number of significant digits. We have supplemented our calculations with a discussion on the subtleties of the boundary conditions of the algebraically special mode. Moving forward, spectral methods should be a very useful tool in finding quasinormal modes on the negative imaginary axis.

Encouraged by these successes, we believe that \texttt{SpectralBP} may serve as a useful tool for the black-hole physics community or just about anyone seeking to solve a differential eigenvalue problem. Future work will look into applications of \texttt{SpectralBP} to the Kerr spacetime, as well as several algorithmic enhancements (such as a novel inverse iteration method) that we shall describe in a later paper. We have also used \texttt{SpectralBP} to discover new interesting properties of the quasinormal modes of Schwarzschild-anti-de Sitter spacetime, which will also be discussed in a later paper.

\begin{appendices}

\section*{Acknowledgements}

We are grateful to Reginald Bernardo and Marc Casals for constructive criticism and many insightful comments on an early version of this paper. We also thank Emanuele Berti, Vitor Cardoso, and Jonathan Thornburg for their encouraging feedback and for pointing us to references that greatly clarified our understanding of the algebraically special modes of Schwazschild. SJCF is supported by the Department of Science and Technology Advanced Science and Technology Human Resources Development Program$-$National Science Consortium. This research is supported by the University of the Philippines Diliman Office of the Vice Chancellor for Research and Development through Project No.~191937 ORG.

\section{Appendix}
\label{sec:appendix}
In this section, we go into further detail of the implementation of Bernstein polynomials into \texttt{SpectralBP}.  Standard references for numerical linear algebra include \cite{TrefethenBau} and \cite{Saad2011}.

\subsection{Closed-form matries}

In Section \ref{sec:pseudospectral}, we derived closed form expressions for converting an operator-function pair $(\hat{f}(u),\phi(u))$ into a matrix-vector pair $(\bm{T}_{j,k},C_k)$ and arrived at equation \eqref{eq:spectralmatrix1} for some generic grid. In \texttt{SpectralBP}, we have implemented using equally spaced and Chebyschev grids.

We insert the definition of Bernstein polynomials in \eqref{eq:BPdefinition}, and simplify factorials containing $N_{\mathrm{max}}$ using the Pocchammer symbol with
\begin{multline}
\dfrac{(N_{\mathrm{max}})!}{(N_{\mathrm{max}}-n)!}\take{N_{\mathrm{max}}-n}{k+q+l-n} \\= \take{N_{\mathrm{max}}}{k+q+l} (k+q+l+1-n)_n.
\end{multline}
In the interest in keeping expressions concise, we define $I^{(n)}(j,k,l,m)$ as the part of our expression that is independent of the grid chosen,
\begin{multline}
I^{(n)}(j,k,l,m) = (-1)^l \take{n}{l}\take{n}{m} \times \\
 \take{N_{\mathrm{max}}}{k+q+l} (k+q+l+1-n)_n
\end{multline}
These manipulations give us
\begin{multline}
\bm{T}_{j,k}^{(n)} = \dfrac{f(u_{j+q})}{(b-a)^{N_{\mathrm{max}}+n}} \sum_{l=0}^n \sum_{m=0}^n I^{(n)}(j,k,l,m) \times \\
 (u_{j+q}-a)^{k+q+l+m-n} \times \\
  (b - u_{j+q})^{N+r+n-k-l-m}.
\end{multline}
We may now plug-in the following equally spaced and Chebyschev grids,
\begin{eqnarray}
u_{j+q}^{\mathrm{equal}} &=& a + (b-a) \dfrac{j+q}{N_{\mathrm{max}}}, \label{eq:collocation1}\\
u_{j+q}^{\mathrm{Cheb}} &=& a + \dfrac{(b-a)}{2} \left[ 1 - \cos \left( \dfrac{j+q}{N_{\mathrm{max}}} \pi \right) \right].\label{eq:collocation2}
\end{eqnarray}
The corresponding matrices simplify to
\begin{multline}
\bm{T}_{j,k}^{\mathrm{equal}} = \dfrac{f(u_{j+q})}{(b-a)^{n}} \sum_{l=0}^n \sum_{m=0}^n I^{(n)}(j,k,l,m) \times \\
(j+q)^{k+q+l+m-n} 
(N+r-j)^{N+r+n-k-l-m},
 \label{eq:implementedspecmatrix1} 
\end{multline}
\begin{multline}
\bm{T}_{j,k}^{\mathrm{Cheb}} = \dfrac{f(u_{j+q})}{(b-a)^{n}} \sum_{l=0}^n \sum_{m=0}^n I^{(n)}(j,k,l,m) \times  \\ 
 \left[1 - \cos \left( \dfrac{j+q}{N_{\mathrm{max}}}\pi \right) \right]^{k+q+l+m-n} \times \\
  \left[1+ \cos \left( \dfrac{j+q}{N_{\mathrm{max}}}\pi \right) \right]^{N+r+n-k-l-m}.
  \label{eq:implementedspecmatrix2}
\end{multline}

\subsection{From GEP to EP}
\label{subsec:GEPtoEP}
Compared to GEPs, the methods for solving eigenvalue problems of the standard form (EPs) are more diverse and more studied. Iterative algorithms to solve either the entire set of eigenvalues and eigenvectors or its subsets are widely available for a general class of complex-valued matrices. Critically, EPs are numerically cheaper to solve than GEPs. 

Consider the polynomial eigenvalue ODE found in Section \ref{subsect:PEP}. If one of the matrices in the GEP is non-singular, then the GEP can be converted into an EP. This is apparently dependent on whether the lowest or highest matrix, $\bm{M}_0$ and $\bm{M}_m$, in the matrix pencil \eqref{eq:polynomialmatrix} are invertible.

The corresponding eigenvalue problems follows,
\begin{equation}
\bm{\mathcal{M}}_1 \bm{\mathcal{C}} = \omega^{-1} \bm{\mathcal{C}}, \qquad
\bm{\mathcal{M}}_2 \bm{\mathcal{C}} = \omega \bm{\mathcal{C}},
\end{equation}
\noindent where
\begin{equation}
\bm{\mathcal{M}}_1 = \left( \begin{array}{cccc}
-\bm{M}_0^{-1} \bm{M}_1 & \dots & -\bm{M}_0^{-1} \bm{M}_{m-1} & -\bm{M}_0^{-1} \bm{M}_{m} \\
\mathds{1} & \dots & 0 & 0 \\
\vdots & \ddots & \vdots & \vdots \\
0 & \dots & \mathds{1} & 0
\end{array} \right),
\end{equation}

\noindent and
\begin{equation}
\bm{\mathcal{M}}_2 = \left( \begin{array}{cccc}
0 & \mathds{1} & \dots & 0 \\
\vdots & \vdots & \ddots & \vdots \\
0 & 0 & \dots & \mathds{1} \\
-\bm{M}_m^{-1} \bm{M}_0 & -\bm{M}_m^{-1} \bm{M}_1 & \dots & -\bm{M}_m^{-1} \bm{M}_{m-1} \\
\end{array} \right).
\end{equation}

As for the full GEP that arises in Section \ref{subsect:fullPEP}, a similar analysis leads to complications. First, it can be shown that $\tilde{\bm{\mathcal{M}}}'$ is always singular. To show this, let us assume that there exists some $\bm{\mathcal{M}}'_{j,k}$ that is invertible. This is to say that, with respect to the matrix pencil from which $\bm{\mathcal{M}}'_{j,k}$ was constructed 
\begin{equation}
(\bm{M}_{j,k,0} + \omega \bm{M}_{j,k,1} + \omega^2 \bm{M}_{j,k,2} + \dots + \omega^m \bm{M}_{j,k,m})\bm{C}_k = 0
\end{equation}

\noindent the matrix $\bm{M}_{j,k,0}$ is invertible. To illustrate that $\tilde{\bm{\mathcal{M}}}'$ is always singular, we rearrange our simultaneous set of ODE's such that $\bm{\mathcal{M}}'_{j,k}$ is now indexed by $\bm{\mathcal{M}}'_{1,1}$, and then we decompose $\tilde{\bm{\mathcal{M}}}'$ as
\begin{equation}
\tilde{\bm{\mathcal{M}}}' = \left( \begin{array}{cc}
\bm{\mathcal{A}}' & \bm{\mathcal{B}}' \\
\bm{\mathcal{C}}' & \bm{\mathcal{D}}'
\end{array} \right)
\end{equation}

\noindent where
\begin{equation}
\begin{array}{c}
\bm{\mathcal{A}}' = \bm{\mathcal{M}}'_{1,1}, \quad \bm{\mathcal{B}}' = \left( \begin{array}{cccc}
\bm{\mathcal{M}}'_{1,2} & \bm{\mathcal{M}}'_{1,3} & \dots & \bm{\mathcal{M}}'_{1,n}
\end{array} \right), \\
\bm{\mathcal{C}}' = \left( \begin{array}{cccc}
\bm{\mathcal{M}}'_{2,1} & \bm{\mathcal{M}}'_{3,1} & \dots & \bm{\mathcal{M}}'_{n,1}
\end{array} \right)^T
\end{array}
\end{equation}

\noindent and
\begin{equation}
\bm{\mathcal{D}}' = \left( \begin{array}{cccc}
\bm{\mathcal{M}}'_{2,2} & \bm{\mathcal{M}}'_{2,3} & \dots & \bm{\mathcal{M}}'_{2,n} \\
\bm{\mathcal{M}}'_{3,2} & \bm{\mathcal{M}}'_{3,3} & \dots & \bm{\mathcal{M}}'_{3,n} \\
\vdots & \vdots & \ddots & \vdots \\
\bm{\mathcal{M}}'_{n,2} & \bm{\mathcal{M}}'_{n,3} & \dots & \bm{\mathcal{M}}'_{n,n}
\end{array} \right).
\end{equation}

The inverse of $\bm{\mathcal{A}}'$ can be shown to be
\begin{equation}
{\bm{\mathcal{A}}'}^{-1} = \left( \begin{array}{cccc}
\bm{M}_0^{-1} & -\bm{M}_0^{-1} \bm{M}_1 & \dots & -\bm{M}_0^{-1} \bm{M}_m \\
0 & \mathds{1} & \dots & 0 \\
\vdots & \vdots & \ddots & \vdots \\
0 & 0 & \dots & \mathds{1}
\end{array} \right)
\end{equation}

We note that each sub-block in $\bm{\mathcal{A}}', \bm{\mathcal{B}}', \bm{\mathcal{C}}', \bm{\mathcal{D}}'$ is of the form
\begin{equation}
\bm{\mathcal{A}}', \bm{\mathcal{B}}', \bm{\mathcal{C}}', \bm{\mathcal{D}}' \sim 
\left( \begin{array}{cccc}
\bm{a}_1 & \bm{a}_2 & \dots & \bm{a}_m \\
0 & \mathds{1} & \dots & 0 \\
\vdots & \vdots & \ddots & \vdots \\
0 & 0 & \dots & \mathds{1} 
\end{array} \right)
\label{eq:genmatrixform}
\end{equation}

\noindent and that the product of any two matrices of this form is also such a matrix. 

For $\tilde{\bm{\mathcal{M}}}'$ to be invertible, the matrix $\bm{\mathcal{D}}' - \bm{\mathcal{C}}' {\bm{\mathcal{A}}'}^{-1} \bm{\mathcal{B}}'$ must not be singular. However, as we have shown, $\bm{\mathcal{D}}'$ and $\bm{\mathcal{C}}'{ \bm{\mathcal{A}}'}^{-1} \bm{\mathcal{B}}'$ are both matrices whose sub-blocks are of the form given in \eqref{eq:genmatrixform}. Thus, the matrix formed by their difference would be singular, as all of the identity matrices cancel out leaving all except $n-1$ rows to vanish.

On the other hand, the inversion of $\tilde{\bm{\mathcal{M}}}''$ is a rather involved calculation best left for computers.

\subsection{Eigenfunction calculation - inverse iteration}
\label{subsect:inviter}

In this section, we describe briefly the inverse iteration method implemented in \texttt{SpectralBP} to calculate the eigenvectors of a matrix pencil. It has the advantage of working on the matrix pencil directly without the need of linearizing the polynomial eigenvalue problem. For a problem involving $n$ dependent functions, a polynomial degree of $m$, and $N$ collocation points, the size of the matrices involved reduce from $(nmN)^2$ to $(nN)^2$.

Suppose $\mu_l$ is some eigenvalue numerically calculated from the GEP in \eqref{eq:fullmatrixproblem2}. That is, for some eigenvalue $\omega_l$ that exactly satisfies \eqref{eq:fullmatrixproblem2},
\begin{equation}
\mu_l = \omega_l + \epsilon, \qquad \epsilon \ll 1.
\end{equation}

The error $\epsilon$ is sourced from finite precision arithmetic, and should be very small. By definition, $\omega_l$ and its corresponding eigenvector $\bm{v}_l$ should also satisfy the polynomial eigenvalue problem without linearization in Section \ref{subsect:PEP},
\begin{equation}
\bm{A}(\omega_l) \bm{v}_l = 0,
\end{equation}

\noindent where
\begin{equation}
\bm{A}(\omega) = \left( \begin{array}{cccc}
\bm{A}_{1,1}(\omega) & \bm{A}_{1,2}(\omega) & \dots & \bm{A}_{1,n}(\omega) \\
\bm{A}_{2,1}(\omega) & \bm{A}_{2,2}(\omega) & \dots & \bm{A}_{2,n}(\omega) \\
\vdots & \vdots & \ddots & \vdots \\
\bm{A}_{n,1}(\omega) & \bm{A}_{n,2}(\omega) & \dots & \bm{A}_{n,n}(\omega) \\
\end{array} \right).
\end{equation}

\noindent and $\bm{A}(\omega)_{j,k}$ comes from the corresponding matrix pencil of the $k$th dependent function of the $j$th equation,
\begin{equation}
\bm{A}_{j,k}(\omega) = \bm{M}_{j,k,0} + \omega \bm{M}_{j,k,1} + \dots + \omega^m \bm{M}_{j,k,m}.
\end{equation}

The inverse iteration algorithm is described in Notebook \ref{alg:inverse iteration}.
\begin{algorithm}[H]
\caption{Inverse iteration}\label{alg:inverse iteration}
\begin{algorithmic}[1]
\State Calculate $\bm{A}(\mu)^{-1}$
\State $v^{(0)} = $  a vector with $|| v^{(0)} ||_{2} = 1$ \Comment{initialize $v^{(0)}$}
\For{$k = 1, 2, 3, \dots, k_{max}$}
\State $w = \bm{A}(\mu)^{-1} v^{(k-1)}$
\State $v^{(k)} = \dfrac{w}{||w||_{\infty}} $ \Comment{normalize}
\If{$|| v^{(k)} - v^{(k-1)} ||_{\infty} \leq \delta$}
\Comment{check convergence}
\State Exit \textbf{for} loop
\EndIf
\EndFor
\State Return $v^{(\mathrm{final})}$
\end{algorithmic}
\end{algorithm}

Its output can be shown to be of the form,
\begin{equation}
v^{(\mathrm{final})} = \dfrac{\bm{v}_l}{||\bm{v}_l||_\infty} + O(\epsilon)
\end{equation}

\noindent The eigenvector $\bm{v}$ can then be split apart into the $n$ eigenfunctions in the Bernstein basis.

The algorithm here is part of a more general inverse iteration algorithm that is useful in the calculation of eigenvalue-eigenvector pairs in polynomial and transcendental eigenvalue problems, which will be the subject of a future work.

It is quite sufficient to calculate eigenfunctions at the same BP order the input eigenvalues were derived from. The error of the eigenfunctions is dominated by the use of a \textit{finite} polynomial basis and not by finite precision arithmetic, as should be apparent in the examples discussed in Section \ref{sec:TISE} and Section \ref{sec:QNM}.

\subsection{Eigenfunction manipulations}
\label{subsec:eigenman}

Suppose we start with an eigenfunction of the form given in \eqref{eq:linearODEsol}. In the interest of brevity, we denote the expanded Bernstein basis order as $N_{\mathrm{max}}=N+q+r$. From the linearity of the problem, eigenfunctions are determined up to a normalization constant. We may choose a normalization constant $A$ so that function $\tilde{\psi}(u)$, given by
\begin{equation}
\tilde{\psi}(u) = A \psi(u) = A \sum_{k=0}^N C_{k+q} B^{N_{\mathrm{max}}}_{k+q}(u),
\end{equation}
satisfies some desirable property. The simplest choice is to either set the coefficient of the leading polynomial expansion at either boundary to 1.
\begin{eqnarray}
A^{-1} &=& C_q, \qquad \to \qquad \lim_{u \to a} \tilde{\psi}_a(u) \approx (u-a)^q \\
A^{-1} &=& C_{N-r}, \; \; \; \to \qquad \lim_{u \to b} \tilde{\psi}_b(u) \approx (b-u)^r.
\label{eq:defaultbounds}
\end{eqnarray}

Consider the following weighted $L^2$-norm, 
\begin{equation}
\int_a^b \lvert\psi(u)\rvert^2 w(u) du = |C|^2, \qquad w(u) = \tilde{A} (u-a)^n (b-u)^m,
\label{eq:fullsquareint}
\end{equation}

\noindent with the condition that $n \geq -2q$ and $m\geq -2r$ so that the integral remains finite. Using properties \eqref{eq:BPint} and \eqref{eq:BPproduct} of the Bernstein basis, the integral \eqref{eq:fullsquareint} can be evaluated to
\begin{equation}
|C| = \tilde{A} \sqrt{ \begin{array}{l}
\sum_{k=0}^{N} \sum_{k'=0}^{N}  \dfrac{A_{k,k'}(b-a)^{m+n+1}}{2N_{\mathrm{max}}+m+n+1}
\end{array}}
\label{eq:squarenorm}
\end{equation}
where
\begin{equation}
A_{k,k'} = \dfrac{\take{N_{\mathrm{max}}}{k}\take{N_{\mathrm{max}}}{k'}}{\take{2N_{\mathrm{max}}+m+n}{n+k+k'}}C_{k+q} C^*_{k'+q}
\end{equation}

A third way to normalize would then be,
\begin{equation}
A^{-1} = \lvert C \rvert, \qquad \int_a^b \lvert \tilde{\psi}_c(u) \rvert^2 w(u) du = 1
\end{equation}

When $w(u) = 1$, the resulting function is normalized such that its $L^2$-norm in the interval $[a,b]$ is unity. The weight function may be utilized to calculate the $L^2$-norm in another set of coordinates. This typically arises when the eigenfunctions are calculated in a compactified set of coordinates.

As an example, consider the the coordinate transformation in \eqref{eq:HOTISE1change} and \eqref{eq:HOTISE2change} in solving the harmonic oscillator. To normalize the eigenfunctions in the uncompactified coordinate system, their respective weights are of the form
\begin{equation}
w_1(u) = (v_1 + 1)^{-1} (1 - v_1)^{-1},\qquad w_2(u) = v_2^{-1} (1-v_2)^{-1}
\end{equation}

One may calculate the square difference between two eigenfunctions in this way. Suppose two eigenfunctions $\psi_1(u)$ and $\psi_2(u)$ calculated from a spectral basis of order $N_1$ and $N_2$ respectively.
\begin{equation}
\psi_1(u) = \sum_{k=q}^{N_1-r} C_k B^{N_1}_k(u), \quad \psi_2(u) = \sum_{k=q}^{N_2-r} C'_k B^{N_2}_k(u)
\end{equation}

Let us say that $N_2 \geq N_1$. We may expand the BP basis order of $\psi_1(u)$ using \eqref{eq:BPraising},
\begin{equation}
\psi_1(u) = \sum_{k=q}^{N_1-r} \sum_{j=0}^{N_2 - N_1} \dfrac{\take{N_1}{k} \take{N_2 - N_1}{j}}{\take{N_2}{k+j}} C_k B^{N_2}_{k+j}(u).
\end{equation}
Thus, we may write the difference between the two eigenfunctions as a new sum of BPs of order $N_2$,
\begin{equation}
\psi_2(u) - \psi_1(u) = \sum_{k=q}^{N_2-r} \tilde{C}_k B_k^{N_2}(u),
\label{eq:difference}
\end{equation}
where
\begin{equation}
\tilde{C}_k = C'_k - \sum_{j=0}^{N_2 - N_1} \dfrac{\take{N_1}{m-j}\take{N_2 - N_1}{j}}{\take{N_2}{m}} C_{m-j}.
\end{equation}
One may then calculate the $L^2$-norm of \eqref{eq:difference} using \eqref{eq:squarenorm}.

With the Bernstein basis, it is also quite easy to rescale our function as in
\begin{equation}
\Psi(u) = \tilde{A} (u-a)^{n'} (b-u)^{m'} \psi(u),
\end{equation}
so that the resulting eigenfunction satisfies different asymptotics at the boundaries of the form
\begin{equation}
\begin{matrix}
\lim_{u \to a} \Phi(u) \sim (u-a)^{q+n'}, \\
 \lim_{u \to b} \Phi(u) \sim (b-u)^{r+m'}.\\
\end{matrix}
\end{equation}
with the condition that $n' \geq -q, m' \geq -r$. This is so that $\Psi(u)$ may still expressed in the Bernstein basis. 

The resulting expression follows from the definition of Bernstein polynomials \eqref{eq:BPdefinition}, that is
\begin{eqnarray}
\Psi(u) = \tilde{A} (b-a)^{n'+m'} \sum_{k=0}^N C'_{k+q} B^{N_{\mathrm{max}}+n'+m'}_{k+q+n'}(u)
\label{eq:rescalefunc1}
\end{eqnarray}

\noindent where
\begin{equation}
C'_{k+q} = C_{k+q} \dfrac{\take{N_{\mathrm{max}}}{k+q}}{\take{N_{\mathrm{max}}+n'+m'}{k+q+n'}}
\label{eq:rescalefunc2}
\end{equation}

\end{appendices}
\bibliographystyle{ieeetr}
\bibliography{template}

\end{document}